\documentclass[aps,prd,11pt,notitlepage,longbibliography,nofootinbib,tightenlines,preprintnumbers,superscriptaddress]{revtex4-1}
\pdfoutput=1

\makeatletter
\renewcommand{\p@subsection}{}
\renewcommand{\p@subsubsection}{}
\makeatother

\usepackage{amsmath,amssymb,amsfonts}
\usepackage{graphicx}
\usepackage{color}
\usepackage{tikz}
\usepackage{dsfont}
\usepackage[pdftex]{hyperref}
\hypersetup{colorlinks=true, linkcolor=darkred, citecolor=blue, linktoc=page}
\definecolor{darkred}{rgb}{0.8,0.1,0.1}

\usepackage{diagbox}

\usepackage{subfigure}

\def\cA{{\cal A}}
\def\cB{{\cal B}}

\def\cF{{\cal F}}
\def\cG{{\cal G}}
\def\cL{{\cal L}}

\def\CC{{\mathds{C}}}
\def\RR{{\mathds{R}}}
\def\ZZ{{\mathds{Z}}}

\DeclareMathOperator{\vol}{vol}
\DeclareMathOperator{\Vol}{Vol}

\DeclareMathOperator{\csch}{csch}
\DeclareMathOperator{\Li}{Li}

\makeatletter\def\l@subsubsection#1#2{}%
\makeatother

\newcommand{\nocontentsline}[3]{}
\newcommand{\tocless}[2]{\bgroup\let\addcontentsline=\nocontentsline#1{#2}\egroup}

\def\Im{\mathop{\rm Im}}

\def\Res{\mathop{\rm Res}}

\newcommand{\pslash}{\ensuremath\diagup\!\!\!\!\!{+}}

\begin{document}

\title{Generalized quotients and holographic duals for 5d S-fold SCFTs}

\author{Fabio Apruzzi}
\email{fabio.apruzzi@pd.infn.it}
\affiliation{Dipartimento di Fisica e Astronomia “Galileo Galilei” Universit\`{a} di Padova, Via Marzolo 8, 35131 Padova, Italy
}
\affiliation{INFN, Sezione di Padova
Via Marzolo 8, 35131 Padova, Italy}
\affiliation{Albert Einstein Center for Fundamental Physics, Institute for Theoretical Physics,\\ University of Bern, Sidlerstrasse 5, CH-3012 Bern, Switzerland}

\author{Oren Bergman}
\email{bergman@physics.technion.ac.il}
\affiliation{Department of Physics, Technion, Israel Institute of Technology, Haifa, 32000, Israel}

\author{Hee-Cheol Kim}
\email{heecheol1@gmail.com}
\affiliation{Department of Physics, Pohang University of Science and Technology, Pohang 37673, Republic of Korea}
\affiliation{Asia Pacific Center for Theoretical Physics (APCTP), Pohang 37673, Republic of Korea}

\author{Christoph F.~Uhlemann} 
\email{uhlemann@maths.ox.ac.uk}

\affiliation{Mathematical Institute, University of Oxford, \\
Andrew-Wiles Building,  Woodstock Road, Oxford, OX2 6GG, UK}
\affiliation{Leinweber Center for Theoretical Physics, Department of Physics	\\
University of Michigan, Ann Arbor, MI 48109-1040, USA}

\preprint{LCTP-22-15}

\begin{abstract}
	$\ZZ_n$ S-folds of 5d SCFTs, including $T_N$, which lead to brane webs with $E_{6,7,8}$ 7-branes were discussed recently in \cite{Acharya:2021jsp,Kim:2021fxx}. We generalize the construction to `fractional quotients', which are based on $\ZZ_n$ actions linking multiple copies of the seed theory and lead to $H_{0,1,2}$ 7-branes. We provide the holographic duals for both classes. This expands the space of explicitly known Type IIB $\rm AdS_6$ solutions by incorporating F-theory 7-branes of type $E_{6,7,8}$ and $H_{0,1,2}$, extending previous constructions for O7-planes. We discuss observables including the free energies and link the results to matrix model descriptions.
\end{abstract}

\maketitle
\tableofcontents
\parskip 1mm

\section{Introduction \& Summary}

The systematic understanding of quantum field theory (QFT) beyond perturbative constructions is a major open problem in theoretical physics. At least in the context of supersymmetric theories, intriguing signs point to a hierarchical structure in which strongly-coupled QFTs in 5 and 6 dimensions, whose existence is predicted by string theory, play a distinguished role. That is, they provide organizing principles in the space of general QFTs and new ways to study lower-dimensional theories. An important part in the string theory toolbox for studying QFT are S-folds. Taking a quotient of a string theory construction for a QFT with respect to a symmetry which involves a duality transformation can lead to new theories \cite{Garcia-Etxebarria:2015wns,Aharony:2016kai,Apruzzi:2020pmv,Heckman:2020svr} and often links parent theories with a gauge theory description to S-fold theories with no known Lagrangian description.

In this work we focus on S-folds of 5d superconformal field theories (SCFTs) engineered by 5-brane webs in Type IIB string theory \cite{Aharony:1997ju,Aharony:1997bh}. Such S-folds were recently studied in \cite{Acharya:2021jsp,Kim:2021fxx} (see also \cite{Tian:2021cif}). They  provide interesting links between theories with gauge theory deformations and theories with no gauge theory deformations.
We will first extend this construction to more general `fractional' quotients and then provide holographic duals for the general class of S-folds, extending the supergravity solutions in \cite{DHoker:2016ujz,DHoker:2016ysh,DHoker:2017mds} to incorporate 7-branes of type $E_{6,7,8}$ and $H_{0,1,2}$. We will discuss a sample of observables which become holographically accessible and connect the results to matrix models, whose construction combines supersymmetric localization for the (Lagrangian) seed theories with prepotentials obtained from the S-fold brane webs.

The general strategy for the S-folds in \cite{Acharya:2021jsp,Kim:2021fxx} was to consider 5-brane webs which are invariant under a $\ZZ_n$ subgroup of the Type IIB $SL(2,\ZZ)$ duality transformations combined with a rotation in the plane of the brane web, and then take a quotient with respect to this discrete group. This produces a deficit angle at the fixed point of the $\ZZ_n$ action on the plane and a monodromy corresponding to a 7-brane.
The four 7-brane configurations arising this way are of type
$D_4$, $E_6$, $E_7$, $E_8$ \cite{Sen:1996vd,Dasgupta:1996ij} and correspond, respectively, to $\ZZ_2$, $\ZZ_3$, $\ZZ_4$, $\ZZ_6$ S-folds.
Similar $\ZZ_n$ actions were used for the 4d S-folds of \cite{Garcia-Etxebarria:2015wns,Aharony:2016kai, Apruzzi:2020pmv,Heckman:2020svr}. We extend this construction by considering multiple copies of the parent brane webs, and defining $\ZZ_n$ actions which combine duality transformations with rotations in the plane and transitions between the different copies of the brane web. The S-folds arising from quotients with respect to these generalized $\ZZ_n$ actions may be described as `fractional quotients' of the original web, and they lead to 7-branes of type $H_0$, $H_1$ and $H_2$ at the fixed points.

The construction of the supergravity solutions associated with the S-folds follows analogously. The Type IIB $\rm AdS_6$ solutions associated with 5-brane webs take the form of a warped product of $\rm AdS_6$ and $\rm S^2$ over a disc $\Sigma$. They are characterized by two locally holomorphic functions $\cA_\pm$ on $\Sigma$. Their differentials $\partial\cA_\pm$ have poles on the boundary of $\Sigma$ which correspond to the external 5-branes in the associated brane junction. The Type IIB $SL(2,\ZZ)$ transformations correspond to $SU(1,1)$ transformations acting on the pair $\cA_\pm$.
For brane junctions with a $\ZZ_n$ symmetry, as described above, the associated supergravity solutions have a $\ZZ_n$ symmetry under a combination of the Type IIB $SL(2,\ZZ)$ transformations with $SL(2,\RR)$ transformations of the disc, which permute the poles of $\partial\cA_\pm$.
We then simply take a quotient of the supergravity solutions with respect to these combined transformations.
This geometrically amounts to cutting the disc $\Sigma$ into slices and keeping one slice, with the edges identified.
The generalization to fractional quotients amounts to keeping multiple slices.
Either construction creates a conical deficit and non-trivial monodromy at the fixed point of the $SL(2,\RR)$ transformation of the disc. The result is the creation of 7-branes of type $E_{6,7,8}$ or $H_{0,1,2}$. This is analogous to the incorporation of O7-planes in \cite{Uhlemann:2019lge}.

The S-fold SCFTs in general do not have gauge theory deformations, even if the parent theories did. This makes holographic duals all the more useful. We initiate a discussion of insights that can be obtained from the holographic duals, including the free energies, central charges, defect operators and universal parts of the operator spectrum in the planar limit. We leave more detailed investigation for the future. In the last part we discuss `effective' matrix models for the S-folds. Since the S-folds do not have gauge theory deformations we can not derive matrix model descriptions for BPS observables from supersymmetric localization. However, an effective prepotential can be defined from the brane webs even in the absence of a gauge theory description. We will combine localization results for the parent theories with these effective prepotentials to construct matrix models, and show that they reproduce the free energies in the planar limit. 

An interesting question is whether the fractional quotient construction can be applied in other dimensions. One may, for example, extend the 4d $\mathcal N=3$ S-fold constructions of \cite{Garcia-Etxebarria:2015wns,Aharony:2016kai}, which are based on D3-branes in $\CC^3/\ZZ_n$, by starting with multiple copies of $\CC^3$ and defining $\ZZ_n$ actions linking them, analogously to the discussion for 5-brane webs here. We plan to investigate the corresponding generalized quotients and whether D3-branes probing them lead to well-defined and perhaps new theories, as well as possible relations to the 4d $\mathcal N=2$ S-folds of \cite{Apruzzi:2020pmv}, separately.

{\bf Outline:} In sec.~\ref{sec:S-fold-review} we fix notation for 7-branes, review the 5d S-fold constructions of \cite{Kim:2021fxx}, and extend them to fractional quotients. In sec.~\ref{sec:duals} we spell out the construction of the holographic duals for a sample of S-fold theories. In sec.~\ref{sec:observables} we discuss a sample of observables which immediately follow from the holographic description. In sec.~\ref{sec:matrix-models} we discuss matrix model descriptions.

\section{F-theory 7-branes and S-folds}\label{sec:S-fold-review}

In this section we first review the S-folds of 5d SCFTs constructed by discrete $\mathbb{Z}_n$ quotients of the associated 5-brane webs in Type IIB string theory. They lead to 7-brane singularities of type $E_6,E_7,E_8$ and were discussed in \cite{Acharya:2021jsp,Kim:2021fxx}. We then propose, by generalizing these S-fold constructions, a new type of discrete quotients of 5-brane webs, leading to 7-brane singularities of type $H_2,H_1,H_0$.

A 5-brane web is a web diagram of $(p,q)$ 5-branes in Type IIB string theory drawn on a 2-plane. We label the 5-branes by their charges $(p,q)$. For a single 5-brane $p$, $q$ are relatively prime integers, otherwise the greatest common divisor is the number of 5-branes. Their orientation in the plane is fixed by their charges; they extend along lines of slope $\tan^{-1}(q/p)$.  In the gravity decoupling limit, the brane web describes the Coulomb branch physics of a 5d SCFT on the common worldvolume of the 5-branes. For example, the dimension of the Coulomb branch is determined by the number of compact faces formed by loops of 5-branes. The $(p,q)$-strings stretched between the 5-branes provide BPS particle states in the 5d SCFT.

We can also introduce 7-branes into the 5-brane web. They are localized at points in the web diagram and 5-branes can terminate on 7-branes with the same $(p,q)$ charge. These 7-branes are magnetic sources for the axio-dilaton scalar $\tau=\chi + ie^{-\phi}$. A 7-brane is therefore accompanied by a branch cut and  the axio-dilaton field $\tau$, when it crosses the branch cut of a $[p,q]$ 7-brane, transforms with $SL(2,\mathbb{Z})$ monodromy matrix $K_{[p,q]}$, which is defined as \cite{Greene:1989ya,DeWolfe:1998eu}
\begin{align}
	K_{[p,q]}=\begin{pmatrix} 1+p q & -p^2\\ q^2 & 1-p q\end{pmatrix}~.
\end{align}
In our notation, an $[r,s]$ 7-brane crossing the branch cut of a $[p,q]$ 7-brane counterclockwise becomes a $K_{[p,q]}[r,s]$ 7-brane.
We name some frequently used 7-branes as
\begin{align}
	\mathbf{A}&=[1,0]~, & \mathbf{B}&=[1,-1]~, & \mathbf{C}&=[1,1]~ ,
\end{align}
and also $\mathbf{\tilde B}=[3,1]$ and $\mathbf{\tilde C}=\mathbf{B}$.

7-branes can collapse to a singularity and, when it happens, the singularity supports a local non-Abelian symmetry which becomes a global symmetry in the 5d theory. Hence, the correct non-Abelian global symmetry of the 5d theory at CFT fixed point can be read off by identifying all possible collapsable 7-brane configurations. These 7-brane configurations are associated to Kodaira singularities in F-theory as studied in \cite{Dasgupta:1996ij,Sen:1996vd}. Using the notation in \cite{DeWolfe:1998eu}, some examples can be written as
\begin{align}
	\mathbf{E_n}&=\mathbf{A}^{n-1}\mathbf{\tilde B}\mathbf{\tilde C}^2~,  & \mathbf{D_4}&=\mathbf{A}^4\mathbf{\tilde B\tilde C}~,
	& \mathbf{H_n}&=\mathbf{A}^{n+1}\mathbf{\tilde C}~,
	\nonumber\\
	\mathbf{E_n}&=\mathbf{A}^{n-1}\mathbf{BCB}~, & 
	\mathbf{D_4}&=\mathbf{A}^4\mathbf{BC}~,&
	\mathbf{H_n}&=\mathbf{A}^{n+1}\mathbf{B}~.
\end{align}

The 7-branes, if introduced, induce a deficit angle in the 10d geometry. The deficit angles at the above 7-brane singularities are $\pi$ for $\mathbf{D_4}$, $2\pi \lbrace \frac{2}{3},\frac{3}{4},\frac{5}{6}\rbrace$ for $\lbrace \mathbf{E_6},\mathbf{E_7},\mathbf{E_8}\rbrace$, and $2\pi\lbrace \frac{1}{6},\frac{1}{4},\frac{1}{3}\rbrace$ for $\lbrace \mathbf{H_0},\mathbf{H_1},\mathbf{H_2}\rbrace$.
The arrangement of 5-branes and 7-branes in a brane web has to respect charge conservation. In particular, the deficit angle of the 7-brane singularity must be consistent with the rest of the brane configuration. 
We will now explain how to realize S-folds and their generalizations in 5-brane webs using these features of the 7-brane singularities.

\subsection{S-folds leading to $E_{6,7,8}$ 7-branes}\label{sec:E-S-folds}

We briefly review the S-fold constructions of 5d SCFTs starting from brane webs with $\mathbb{Z}_n$ symmetry,  investigated in \cite{Acharya:2021jsp,Kim:2021fxx}. Consider a web of 5-branes with a $\mathbb{Z}_n$ symmetry generated by a rotation of the $(p,q)$-plane by an angle $2\pi/n$ together with a proper $SL(2,Z)$ transformation of the 5-brane charges. We can then perform a $\mathbb{Z}_n$ quotient of the brane web with respect to this $\mathbb{Z}_n$ symmetry. This quotient can be implemented in two steps. First, we split the brane web into $n$-slices of the same shape, and then we take one slice. The second step is to insert a 7-brane singularity of type $E_6, E_7, E_8$ for $n=3,4,6$, respectively, at the $\mathbb{Z}_n$ fixed point, such that the monodromy introduced by the 7-brane is consistent with the 5-brane configuration in the slice of the brane web we have taken. This quotient is called S-fold of the 5d SCFT since it is a generalization of the $\mathbb{Z}_2$ quotient of brane webs by means of an orientifold with 4 $D7$-branes (or a $D_4$ 7-brane singularity) introduced at the $\mathbb{Z}_2$ fixed point. See for example \cite{Kim:2021fxx} for more details.

\begin{figure}[t]
	\includegraphics[height=40mm]{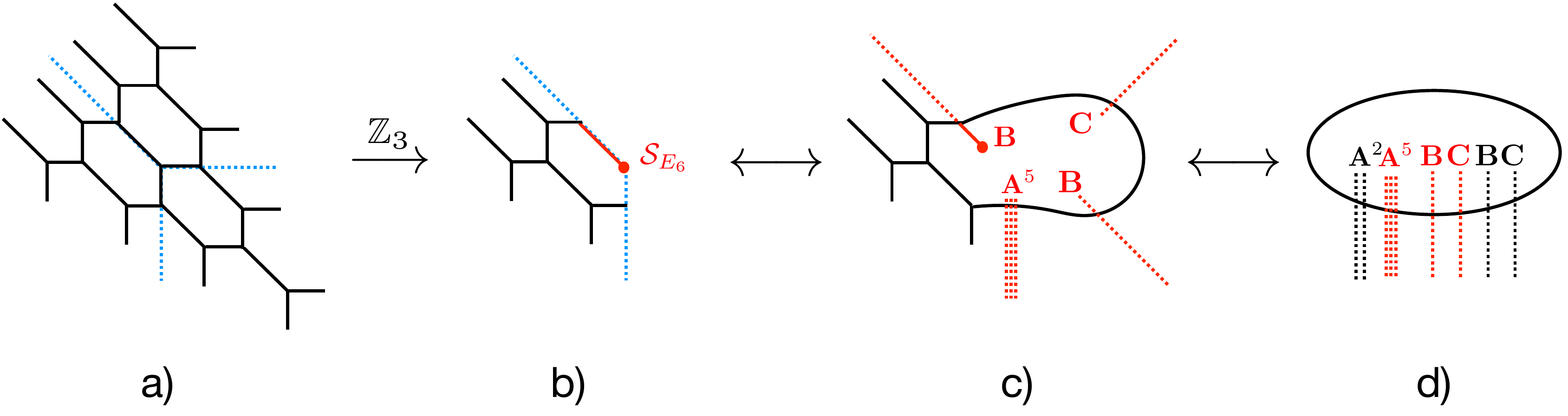}
	\caption{$\mathbb{Z}_3$ quotient of the $T_4$ theory. In b), we took one slice out of 3 slices and added an $E_6$ 7-brane at the fixed point. The 7-brane singularity can be resolved as in c). We then brought all 7-branes inside the 5-brane loop using HW transitions. The final configuration in d) is the brane web for the $SU(2)+7{\bf F}$ theory.}
	\label{fig:T4}
\end{figure}

For instance, the $\mathbb{Z}_3$ quotient of the $T_4$ theory, discussed in \cite{Kim:2021fxx}, is illustrated in fig. \ref{fig:T4}.
The brane web for $T_4$, shown in a), has a $\mathbb{Z}_3$ symmetry. As shown in b), we perform the $\mathbb{Z}_3$ quotient of the brane web by taking the lower left slice and inserting an $E_6$ type 7-brane denoted by $\mathcal{S}_{E_6}$ at the $\mathbb{Z}_3$ fixed point. Note that the fixed point is located at a junction of three 5-branes and, after the quotient, only one of the three 5-branes denoted by a red line is kept and attached to the $E_6$ 7-brane. In c), we have resolved the $E_6$ 7-brane into ${\bf A^5BCB}$ and then, to be consistent with the rest of the 5-brane web and monodromies, the 5-brane must terminate at the ${\bf B}$ brane from the 7-brane singularity. The final brane configuration in d) was obtained by a series of Hanany-Witten (HW) transitions which bring all the external 7-branes inside the loop of 5-branes. The brane web corresponds to the $SU(2)$ gauge theory with 7 fundamental hypermultiplets, which we denote by $SU(2)+7{\bf F}$. This theory has an enhanced $E_8$ flavor symmetry at the UV fixed point which can be seen from the 7-brane configuration ${\bf A^7BCBC}$ inside the 5-brane loop.

\begin{figure}[t]
	\includegraphics[height=55mm]{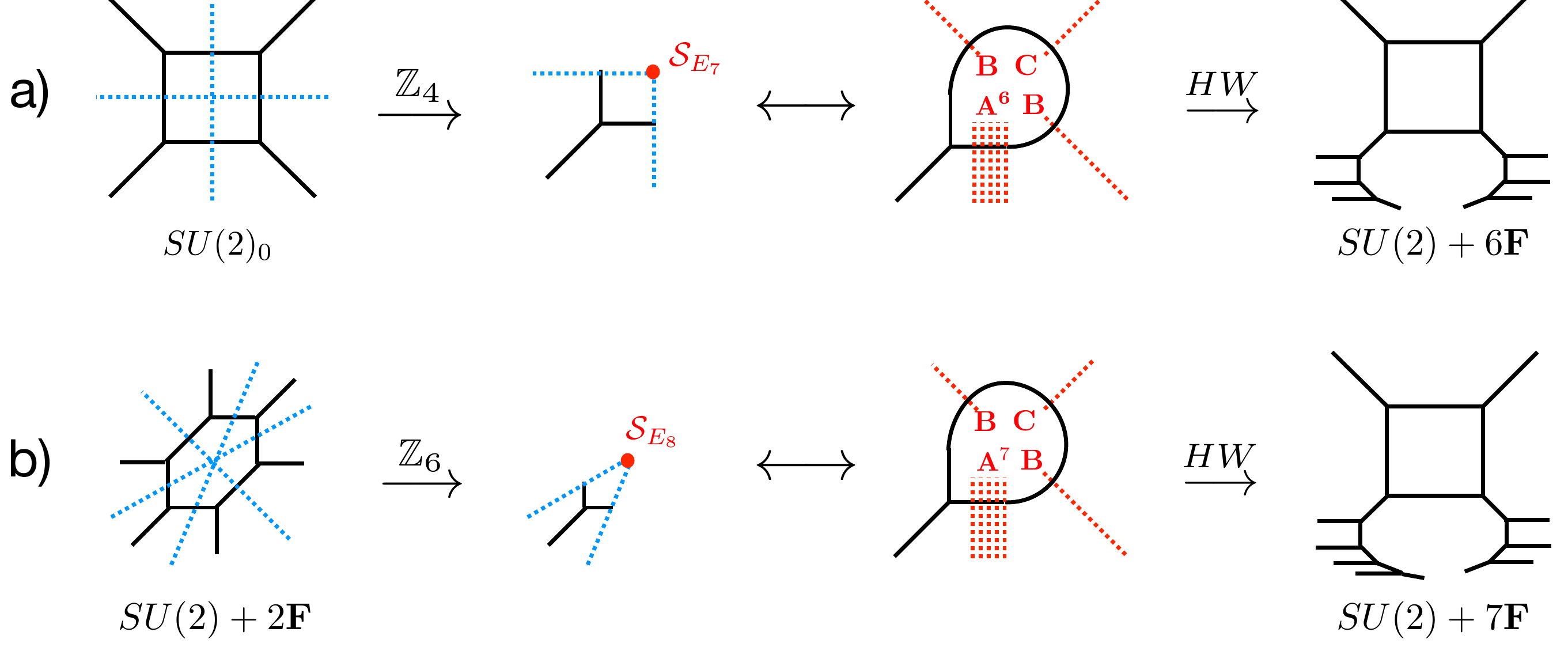}
	\caption{a) $\mathbb{Z}_4$ quotient of the $SU(2)_0$ theory which leads to the $SU(2)+6{\bf F}$ theory. b) $\mathbb{Z}_6$ quotient of the $SU(2)+2{\bf F}$ theory which leads to the $SU(2)+7{\bf F}$ theory.}
	\label{fig:Z4-Z6}
\end{figure}

Examples of $\mathbb{Z}_4$ and $\mathbb{Z}_6$ quotients of 5d SCFTs are presented in fig. \ref{fig:Z4-Z6}. In a), we start from the $SU(2)$ gauge theory at $\theta=0$ with a $\mathbb{Z}_4$ symmetry under $\pi/2$ rotation, and then we perform a $\mathbb{Z}_4$ quotient with a 7-brane singularity of $E_7$ type which can be resolved into ${\bf A^6BCB}$ branes. The resulting diagram at the right end corresponds to the $SU(2)$ gauge theory with 6 fundamental hypermultiplets. In b), a $\mathbb{Z}_6$ quotient of the $SU(2)$ gauge theory with 2 fundamentals is realized by a 7-brane singularity of $E_8$ type which can now be resolved into ${\bf A^7BCB}$ branes. One then ends up with the $SU(2)$ gauge theory with 7 fundamental hypers for the last brane diagram.

These S-fold constructions can be verified by computing the prepotentials for the SCFTs before and after the discrete quotients. How to compute the prepotentials with the S-folds was proposed in \cite{Kim:2021fxx}. The main idea is that the total area of compact faces in the 5-brane web after the $\mathbb{Z}_n$ quotient is reduced to $1/n$ of the total area of the original brane web, and hence the prepotential of the associated 5d SCFT is also reduced by $1/n$ factor. This is because the area of a compact face is given by the first derivative of the prepotential with respect to the K\"ahler parameter dual to the face. Moreover, the K\"ahler parameters in an orbit of the $\mathbb{Z}_n$ symmetry will be identified under the $\mathbb{Z}_n$ quotient. Consequently, the prepotential of the quotient theory will be $1/n$ of the original prepotential with K\"ahler parameters in each $\mathbb{Z}_n$ orbit identified. Thus, we have
\begin{align}
	\mathcal{F}_{\mathcal{T}/\mathbb{Z}_n} = \frac{1}{n}\left.\mathcal{F}_{\mathcal{T}}\right|_{\phi_{S(i)}=\phi_i} \ ,
\end{align}
where $\mathcal{T}$ denotes the original theory and $\mathcal{T}/\mathbb{Z}_n$ denotes the  $\mathbb{Z}_n$ S-fold theory, and $S(i)$ stands for permutations of the K\"ahler parameters in a $\mathbb{Z}_n$ orbit including $\phi_i$.

For example, the $T_4$ theory with brane web in fig. \ref{fig:T4} a) has the prepotential
\begin{align}
	6\mathcal{F}_{T_4} = \sum_{i=1}^3 5\phi_i^3 -3\sum_{i<j}^3(\phi_i^2\phi_j+\phi_i\phi_j^2)+6\phi_1\phi_2\phi_3 \ ,
\end{align}
where we switched off all mass parameters. Here $\phi_i$ denotes the dynamical K\"ahler parameter for each compact face in the brane web. The $\mathbb{Z}_3$ symmetry permutes these three K\"ahler parameters by $S : (\phi_1,\phi_2,\phi_3)\rightarrow (\phi_2,\phi_3,\phi_1)$. The prescription above tells us that the prepotential of the theory after the $\mathbb{Z}_3$ quotient becomes
\begin{align}
	6\mathcal{F}_{T_4/\mathbb{Z}_3} = \frac{1}{3}\times \left.6\mathcal{F}_{T_4}\right|_{\phi_1=\phi_2=\phi_3} = \phi_1^3 \ .
\end{align}
Indeed, this agrees precisely with the prepotential of the $SU(2)$ theory with 7 fundamental hypermultiplets, as expected from the S-fold construction using the brane web in fig. \ref{fig:T4} d).

One can similarly calculate the prepotentials of the 5d SCFTs with $\mathbb{Z}_4,\mathbb{Z}_6$ S-folds in fig. \ref{fig:Z4-Z6} and check them against the predictions from the brane webs. The prepotential after the $\mathbb{Z}_4$ quotient of the $SU(2)_0$ theory at the right end of fig.  \ref{fig:Z4-Z6} a) is given by
\begin{align}
	6\mathcal{F}_{SU(2)_0} = 8\phi^3 \quad \overset{\mathbb{Z}_4}{\longrightarrow} \quad 6\mathcal{F}_{SU(2)_0/\mathbb{Z}_4} = \frac{1}{4}\times 6\mathcal{F}_{SU(2)_0} = 2\phi^3 \ ,
\end{align}
which coincides with the prepotential of the $SU(2)$ gauge theory with 6 fundamentals as predicted from the brane web. Similarly, the prepotential of the $\mathbb{Z}_6$ quotient acting on the $SU(2)$ theory with 2 fundamentals is
\begin{align}
	6\mathcal{F}_{SU(2)+2{\bf F}} = 6\phi^3 \quad \overset{\mathbb{Z}_6}{\longrightarrow} \quad 6\mathcal{F}_{SU(2)+2{\bf F}/\mathbb{Z}_6} = \frac{1}{6}\times 6\mathcal{F}_{SU(2)+2{\bf F}} = \phi^3 \ .
\end{align}
This perfectly agrees with the prepotential of the $SU(2)$ gauge theory with 7 fundamental hypers in fig. \ref{fig:Z4-Z6} b).

\subsection{S-folds leading to $H_{0,1,2}$ 7-branes}\label{sec:H-S-folds}
In this section, we propose a new operation on brane webs with $\mathbb{Z}_3,\mathbb{Z}_4,\mathbb{Z}_6$ symmetry, using 7-branes of type $H_2,H_1,H_0$, respectively. The operation is a generalization of the S-folds discussed in the previous subsection. Since the deficit angles for these 7-branes are less than $\pi$, we cannot obtain them by $\mathbb{Z}_n$ quotients of a single brane web. Instead, we suggest the following operation.

First, we cut the 5-brane web with $\mathbb{Z}_n$ symmetry into $n$ slices, as before. Then we remove one slice, and introduce a 7-brane of type $H_2,H_1,H_0$ for $n=3,4,6$, respectively, at the center. If the center is located on a 5-brane junction, which can only happen for $n=3,4$, we keep $n-1$ 5-branes meeting at the junction and attach them to the 7-brane we have introduced.
Each 5-brane, when the 7-brane at the center is resolved into a collection of 7-branes, will anchor on one of the 7-branes with the same charge. The final configuration, having $n-1$ slices of the original brane web, possibly with $n-1$ 5-branes ending on the 7-brane at the center, is consistent with the deficit angle of the 7-brane\footnote{Note that when we resolve the 7-brane singularity at the center as $H_0={\bf AB}, H_1={\bf A^2B}, H_2={\bf A^3B}$, we may first need to act with an $SL(2,\mathbb{Z})$ transformation on the brane web so that the 5-brane charges become compatible with the resolved 7-brane charges.}. This operation realizes an ``$(n-1)/n$ quotient'' of the brane web with $\mathbb{Z}_n$ symmetry. In fact, this can also be understood as a $\mathbb{Z}_n$ quotient of $n-1$ copies of the same 5-brane web, which will be explained in more detail below.

\begin{figure}[t]
	\includegraphics[height=35mm]{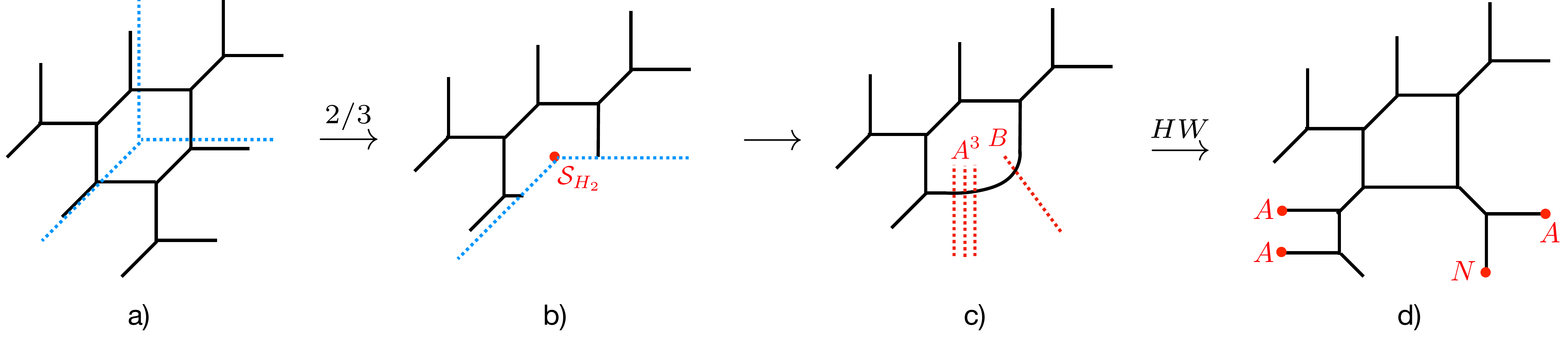}
	\caption{$2/3$ quotient of the $T_3$ theory. The resulting theory is $SU(2)+6{\bf F}$. The red dots in d) are the 7-branes from the $H_2$ singularity introduced at the center of the quotient. We used the relation ${\bf AB=NA}$.}
	\label{fig:T3-2/3}
\end{figure}

For example, the $2/3$ quotient of the $T_3$ theory is illustrated in fig. \ref{fig:T3-2/3}. In b), we keep $2$ slices of the $T_3$ diagram as a physical domain of the brane web and introduce a 7-brane of type $H_2$ at the fixed point of the $\mathbb{Z}_3$ rotation. The $H_2$ singularity can be resolved by ${\bf A^3B}$ 7-branes as drawn in c) and a series of Hanany-Witten (HW) transitions eventually leads to the last diagram d) which represents a 5-brane web for the $SU(2)$ gauge theory with 6 fundamental hypermultiplets. Therefore, we claim that the $2/3$ quotient of the $T_3$ theory gives rise to the $SU(2)+6{\bf F}$ theory. 

\begin{figure}[t]
	\includegraphics[height=65mm]{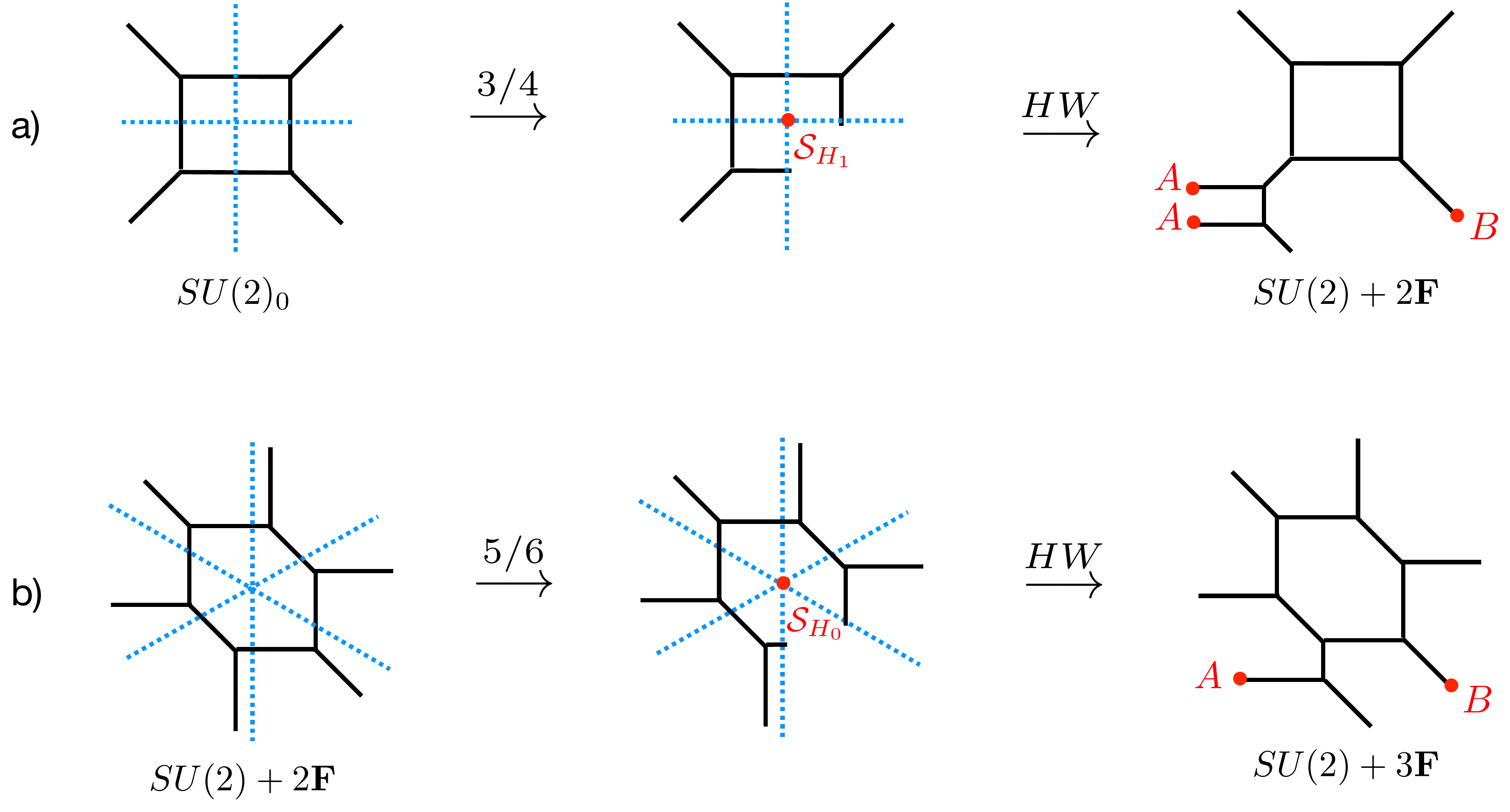}
	\caption{a) $3/4$ quotient of the $SU(2)_0$ theory realized by an $H_1$ 7-brane. This gives rise to the $SU(2)+2{\bf F}$ theory. b) $5/6$ quotient of the $SU(2)+2{\bf F}$ theory realized by an $H_0$ 7-brane, which leads to the $SU(2)+3{\bf F}$ theory. The red dots denote the 7-branes coming from the $H_n$ 7-branes at the center.}
	\label{fig:ex-3/4-5/6}
\end{figure}

Similarly, the $3/4$ quotient of the $SU(2)$ gauge theory at $\theta=0$ can be performed by keeping $3$ slices out of $4$ slices in the original diagram, leading to an $H_1$ 7-brane at the center as shown in fig. \ref{fig:ex-3/4-5/6} a). The result is the $SU(2)$ gauge theory with 2 fundamental hypers. The $5/6$ quotient of the $SU(2)$ gauge theory with 2 fundamentals is depicted in fig. \ref{fig:ex-3/4-5/6} b), which leads to the $SU(2)$ theory with 3 fundamentals.

\begin{figure}[t]
	\includegraphics[height=28mm]{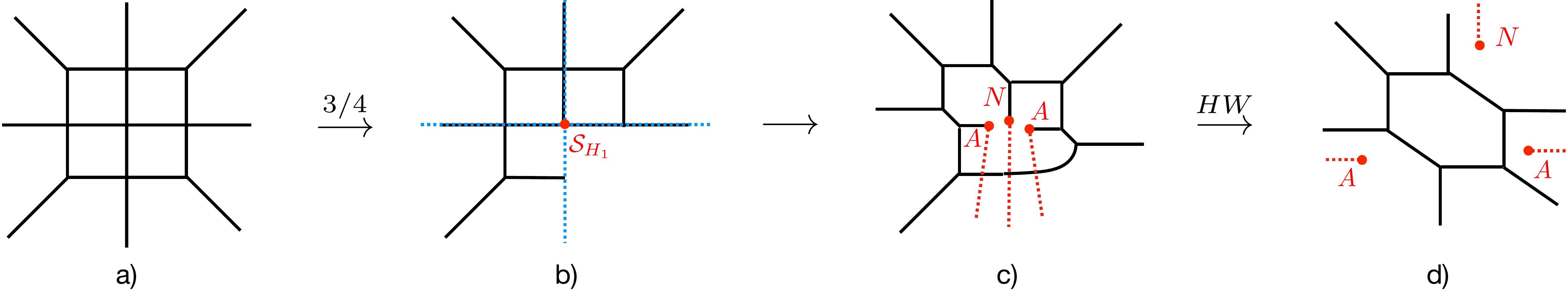}
	\caption{a) $3/4$ quotient of the $[1]-SU(3)_0-SU(3)_0-[1]$ quiver gauge theory. The resulting theory in d) is the $SU(2)+2{\bf F}$ theory. In c), we used ${\bf H_1} = {\bf A^2B}={\bf ANA}$ and we brought all these 7-branes outside the compact face along their charge directions in d).}
	\label{fig:ex-3/4}
\end{figure}

Another interesting example is the $3/4$ quotient of the $SU(3)_0\times SU(3)_0$ quiver gauge theory with 1 fundamental hyper for each node, where the subscript of a gauge group denotes the Chern-Simons level. This is illustrated in fig. \ref{fig:ex-3/4}. In this case, the $\mathbb{Z}_4$ fixed point is located on the junction of four 5-branes. Under the $3/4$ quotient, we keep only three 5-branes at the junction and attach them to the $H_1$ singularity at the fixed point, as drawn in b). The $H_1$ 7-brane can be resolved into ${\bf A^2B}={\bf ANA}$ 7-branes with 5-branes attached as shown in c). These 7-branes can be moved out of the web via HW-transitions, eliminating the attached 5-branes. They can then finally be removed by bringing them to infinity as shown in d). We obtain the $SU(2)$ gauge theory with 2 fundamentals at the end.

We shall now interpret the $(n-1)/n$ quotients of a single brane web discussed so far as a $\mathbb{Z}_n$ quotient acting on $n-1$ identical copies of the brane web. Consider a brane web with $\mathbb{Z}_n$ symmetry. We prepare $n-1$ copies of this brane web and draw them on  separate planes. We decompose each brane web into $n$ slices, as before. The slices of the first web are labeled by $k=0,\ldots, n-1$, those of the second by $k=n,\ldots,2n-1$, and accordingly for the remaining copies. We then define a new $\ZZ_n$ action as a reshuffling of the slices according to $k\rightarrow k+n-1 \mod n(n-1)$. This can be understood as a rotation of all planes by the same $2\pi\times \frac{n-1}{n}$ angle, with the refinement that each slice crossing the dividing line between the first and last slice of a sheet gets bumped to the next sheet.
This is indeed an operation of  degree $n$: applying it $n$ times is an identity. This $\mathbb{Z}_n$ action permutes the compact faces in the $n-1$ brane webs and exchanges the K\"ahler parameters associated to each face as
\begin{align}
	\mathbb{Z}_n \ : \	\phi_i^{(a)} \ \rightarrow  \ \phi_{S(i)}^{(a')} \ ,
\end{align}
where $\phi^{(a)}_i$ denotes the K\"ahler parameter for the $i$-th surface on the $a$-th sheet with $\phi_i^{(n)}=\phi^{(1)}_i$ and $S(i)$ stands for the location of the $i$-th surface after a $2\pi\times \frac{n-1}{n}$ rotation. If the $\mathbb{Z}_n$ action brings the $i$-th face to the next sheet $a'=a+1$, otherwise $a'=a$.

We now take a quotient of this union of brane webs with respect to the new $\mathbb{Z}_n$ symmetry. The quotient identifies all compact faces and 5-branes in a $\mathbb{Z}_n$ orbit. This also identifies the K\"ahler parameters in the orbit, i.e. $\phi_i^{(a)} =\phi_{S(i)}^{(a')}$. We finally keep only one representative for each $\mathbb{Z}_n$ orbit. The resulting brane diagram amounts to keeping $n-1$ slices out of the $n$ slices in the original brane web. By an abuse of notation we call this $\mathbb{Z}_n$ quotient of $n-1$ copies of a theory $\mathcal{T}$, $(\mathcal{T})^{n-1}/\mathbb{Z}_n$, or a fractional quotient $(\mathcal T)^{(n-1)/n}$.

When the $\mathbb{Z}_n$ fixed point is located at the center of a face
the rank of the $\mathbb{Z}_n$ folded theory becomes
\begin{align}
	r\left((\mathcal{T})^{n-1}/\mathbb{Z}_n\right) = \frac{n-1}{n}(r(\mathcal{T})-1)+1 \ ,
\end{align}
where $r(\mathcal{T})$ is the rank of the seed theory $\mathcal{T}$.
When the $\mathbb{Z}_n$ fixed point is located on a vertex of 5-branes the rank becomes
\begin{align}
	r\left((\mathcal{T})^{n-1}/\mathbb{Z}_n\right) = \frac{n-1}{n}(r(\mathcal{T})-n)+1 \ .
\end{align}
In the latter case, `$-n$' is for the $n$ faces adjacent to the 5-brane junction at the center, and we added `1' (instead of `$n-1$') since the 5-brane junction will be resolved into  $n-1$ seperate 5-branes and these $n$ faces in a $\mathbb{Z}_n$ orbit will be all identified after the $\mathbb{Z}_n$ quotient.

\begin{figure}[t]
	\includegraphics[height=43mm]{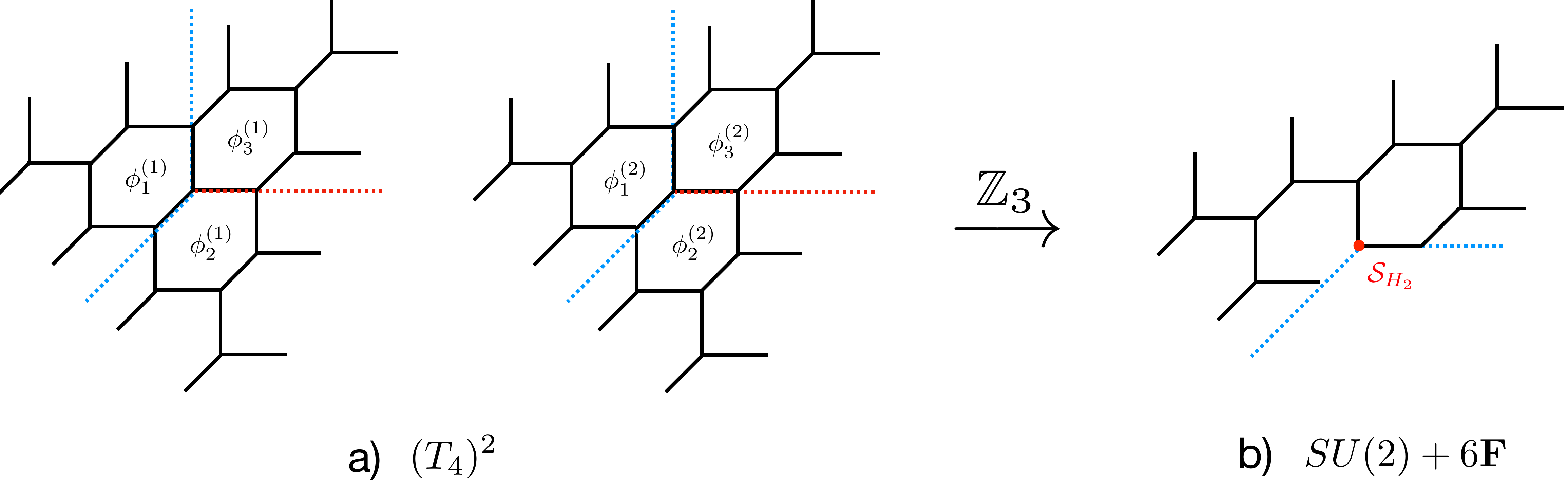}
	\caption{$\mathbb{Z}_3$ quotient of two copies of the $T_4$ theory. In a), we have two $T_4$ diagrams connected through branch cuts denoted by red dotted lines. The resulting theory in b) is the $SU(2)+6{\bf F}$ theory.}
	\label{fig:T4-2-3}
\end{figure}

For example, in fig. \ref{fig:T4-2-3}, we present a $\mathbb{Z}_3$ folding of two copies of the $T_4$ theory. Here, the $\mathbb{Z}_3$ action is a $\frac{4\pi}{3}$ rotation of two sheets which permutes the K\"ahler parameters as $\phi^{(1)}_1\rightarrow\phi^{(2)}_3\rightarrow \phi^{(2)}_2$ and $\phi^{(1)}_2\rightarrow\phi^{(2)}_1\rightarrow \phi^{(1)}_3$. The $\mathbb{Z}_3$ quotient identifies the faces in the $\mathbb{Z}_3$ orbits and thus identifies the K\"ahler parameters as $\phi^{(1)}_1=\phi^{(2)}_3= \phi^{(2)}_2$ and $\phi^{(1)}_2=\phi^{(2)}_1= \phi^{(1)}_3$. Then we insert a 7-brane of type $H_2$ at the fixed point and obtain the diagram in b). Note that since the 5-branes at the junction terminate on distinct 7-branes when the $H_2$ 7-brane at the singularity is resolved, two compact faces in b) are actually identified, and thus $\phi^{(1)}_1=\phi^{(1)}_3$. Therefore, the $\mathbb{Z}_3$ quotient of two copies of the $T_4$ theory gives rise to the brane web in b) which represents the $SU(2)$ gauge theory with 6 fundamentals.

This interpretation of the $(n-1)/n$ quotient using multiple sheets of 5-brane webs turns out to be quite useful for calculating the prepotential of the  theories after the quotient. We start with $n-1$ copies of the prepotential for the original 5-brane web. The $\mathbb{Z}_n$ quotient can then be implemented by deviding it by $n$ and identifying the K\"ahler parameters in each $\mathbb{Z}_n$ orbit as guided by the $\mathbb{Z}_n$ quotient defined above. This leads to the prepotential of the theory after the $(n-1)/n$ quotient, which we call $(\mathcal{T})^{n-1}/\mathbb{Z}_n$, as 
\begin{align}\label{eq:F-fractional}
	\mathcal{F}\left((\mathcal{T})^{n-1}/\mathbb{Z}_n\right) = \frac{1}{n} \sum_{a=1}^{n-1}\left.\mathcal{F}(\mathcal{T}(\phi^{(a)}))\right|_{\phi^{(a')}_{S(i)}=\phi^{(a)}_i} \ .
\end{align}
We remark that the K\"ahler parameters for the surfaces around the 5-brane junction at the center are all identified.

We can verify the prepotentials obtained this way against those of the theories identified from the $(n-1)/n$ quotient of brane webs in the examples above. For instance, the prepotentials for the theories in fig. \ref{fig:T3-2/3} and fig. \ref{fig:ex-3/4-5/6} can be computed as
\begin{align}
	6\mathcal{F}_{T_3} = 3\phi^3 \quad \overset{2/3}{\longrightarrow} \quad  &  \frac{1}{3}\left(3(\phi^{(1)})^3+3(\phi^{(2)})^3\right) = 2\phi^3 =6\mathcal{F}_{SU(2)+6{\bf F}} \ , \nonumber \\
	6\mathcal{F}_{SU(2)_0} = 8\phi^3 \quad \overset{3/4}{\longrightarrow} \quad  & \frac{1}{4}\sum_{a=1}^3 8\left.(\phi^{(a)})^3\right|_{\phi^{(a)}=\phi} = 6\phi^3 = 6\mathcal{F}_{SU(2)+2{\bf F}} \ , \nonumber \\
	6\mathcal{F}_{SU(2)+2{\bf F}} = 6\phi^3 \quad \overset{5/6}{\longrightarrow} \quad  & \frac{1}{6}\sum_{a=1}^5 6\left.(\phi^{(a)})^3\right|_{\phi^{(a)}=\phi} = 5\phi^3 = 6\mathcal{F}_{SU(2)+3{\bf F}} \ ,
\end{align}
which agree with the final brane webs after the quotient with $H_2,H_1,H_0$ 7-branes. Also the $T_4$ quiver theory in fig. \ref{fig:T4-2-3} after the $2/3$ quotient becomes a theory with the prepotential
\begin{align}
	&6\mathcal{F}_{T_4} = \sum_{i=1}^3 5\phi_i^3 -3\sum_{i<j}^3(\phi_i^2\phi_j+\phi_j^2\phi_i)+6\phi_1\phi_2\phi_3 \nonumber \\
	& \overset{2/3}{\longrightarrow} \quad  \frac{1}{3}\left.\left(6\mathcal{F}_{T_4}(\phi^{(1)})+6\mathcal{F}_{T_4}(\phi^{(2)})\right)\right|_{\phi^{(a)}_i=\phi} = 2\phi^3 = 6\mathcal{F}_{SU(2)+6{\bf F}} \ .
\end{align}
This is indeed in agreement with the prepotential of the $SU(2)$ gauge theory with 6 fundamentals as predicted from the brane web.

\begin{figure}[t]
	\includegraphics[height=43mm]{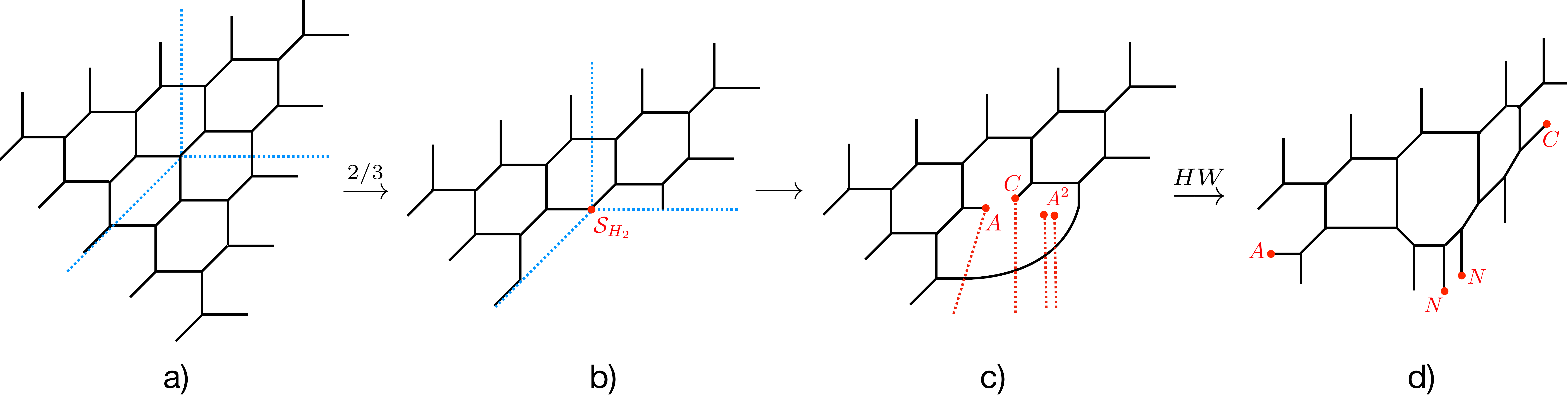}
	\caption{$\mathbb{Z}_3$ quotient of $T_5$ theory. The resulting theory in d) is the $SU(4)_0+10{\bf F}$ theory. We used relations $H_2={\bf A^3B} = {\bf ACA^2}$. The red dots in d) are the 7-branes arising from the $H_2$ singularity.}
	\label{fig:T5-2-3}
\end{figure}

The last example in this subsection is the $2/3$ quotient of the $T_5$ theory. We illustrate this in fig. \ref{fig:T5-2-3}. After the quotient with a 7-brane singularity of type $H_2$, we obtain the $SU(4)_0$ gauge theory with 10 fundamental hypers. This can be verified by computing the prepotentials before and after the quotient as follows. The prepotential of the $T_5$ theory is given by
\begin{align}
	6\mathcal{F}_{T_5}=&\sum_{i=1}^35\phi_i^3 +\sum_{i=4}^66\phi_i^3-3\phi_1\phi_4(\phi_1+\phi_4)-3\phi_1\phi_6(\phi_1+\phi_6)-3\phi_2\phi_4(\phi_2+\phi_4)-3\phi_2\phi_5(\phi_2+\phi_5) \nonumber \\
	&-3\phi_3\phi_5(\phi_3+\phi_5)-3\phi_3\phi_6(\phi_3+\phi_6)-3\phi_4\phi_5(\phi_4+\phi_5)-3\phi_4\phi_6(\phi_4+\phi_6)-3\phi_5\phi_6(\phi_5+\phi_6) \nonumber \\&+6\phi_1\phi_4\phi_6+6\phi_2\phi_4\phi_5+6\phi_3\phi_5\phi_6+6\phi_4\phi_5\phi_6 \ .
\end{align}
The $\mathbb{Z}_3$ action on two copies of the $T_5$ theory can be realized by the permutation of the K\"ahler parameters given by
\begin{align}
	\phi^{(1)}_1 \rightarrow \phi^{(2)}_3 \rightarrow \phi^{(2)}_2 \ , \quad \phi^{(1)}_2 \rightarrow \phi^{(2)}_1 \rightarrow \phi^{(1)}_3 \ , \quad \phi^{(1)}_4 \rightarrow \phi^{(1)}_6 \rightarrow \phi^{(2)}_5 \ , \quad \phi^{(1)}_5 \rightarrow \phi^{(2)}_4 \rightarrow \phi^{(2)}_6 \ .
\end{align}
After the $\mathbb{Z}_3$ quotient, these parameters are identified as 
\begin{align}
	\phi_1 \equiv \phi^{(1)}_1=\phi^{(2)}_3=\phi^{(2)}_2 \ , \quad \phi_2 \equiv \phi^{(1)}_2=\phi^{(2)}_1=\phi^{(1)}_3 \ , \quad \phi_3 \equiv \phi^{(1)}_{4,5,6}=\phi^{(2)}_{4,5,6}\ .
\end{align}
Thus we obtain the prepotential for the $\mathbb{Z}_3$ folded theory $(T_5)^2/\mathbb{Z}_3$, after these identifications of the K\"ahler parameters,
\begin{align}
	6\mathcal{F}_{(T_5)^2/\mathbb{Z}_3} = \frac{1}{3}\left(6\mathcal{F}_{T_5}(\phi^{(1)})+6\mathcal{F}_{T_5}(\phi^{(2)})\right) = 5\phi_1^3+5\phi_2^3+4\phi_3^3-6\phi_4(\phi_1^2+\phi_2^2) = 6\mathcal{F}_{SU(4)_0+10{\bf F}} \ .
\end{align}
This coincides with the prepotential of the $SU(4)_0$ theory with 10 fundamentals as expected.

\section{S-folds and F-theory 7-branes in A\lowercase{d}S$_6$/CFT$_5$}\label{sec:duals}

In this section we construct holographic duals for $\ZZ_3$ quotients of the $T_N$ theories, $\ZZ_4$ quotients of the $+_{N,N}$ theories and $\ZZ_6$ quotients of the $\pslash_N$ theories of \cite{Bergman:2018hin}, as well as the 2/3, 3/4 and 5/6 quotients leading to $H_{0,1,2}$ 7-branes.
The construction principle is general, and can be adapted to other theories straightforwardly.
We briefly review the general AdS$_6$ solutions in Type IIB first, emphasizing the relevant $SL(2,\ZZ)$ transformations, and then proceed to discussing the quotients.

The $\rm AdS_6$ solutions \cite{DHoker:2016ujz,DHoker:2016ysh,DHoker:2017mds} are specified by two locally holomorphic functions $\cA_\pm$ on a Riemann surface $\Sigma$. The geometry is a warped product of $\rm AdS_6$ and $\rm S^2$ over $\Sigma$.
The metric in Einstein frame, axion-dilaton scalar $B=(1+i\tau)/(1-i\tau)$ and complex two-form $C_{(2)}$ are given by
\begin{align}\label{eqn:ansatz}
	ds^2 &= f_6^2 \, ds^2 _{\mathrm{AdS}_6} + f_2^2 \, ds^2 _{\mathrm{S}^2} 
	+ 4\rho^2\, |dw|^2~,
	\qquad\quad
	B =\frac{\partial_w \cA_+ \,  \partial_{\bar w} \cG - R \, \partial_{\bar w} \bar \cA_-   \partial_w \cG}{
		R \, \partial_{\bar w}  \bar \cA_+ \partial_w \cG - \partial_w \cA_- \partial_{\bar w}  \cG}~,
	\nonumber\\
	C_{(2)}&=\frac{2i}{3}\left(
	\frac{\partial_{\bar w}\cG\partial_w\cA_++\partial_w \cG \partial_{\bar w}\bar\cA_-}{3\kappa^{2}T^2} - \bar{\mathcal{A}}_{-} - \mathcal{A}_{+}  \right)\vol_{S^2}~.
\end{align}
$ds^2_{\rm AdS_6}$ and $ds^2_{\rm S^2}$ are the line elements for unit-radius $\rm AdS_6$ and $\rm S^2$, respectively, and $w$ is a complex coordinate on $\Sigma$.
The metric functions are given by
\begin{align}\label{eq:metric-functions}
	f_6^2&=\sqrt{6\cG T}~, & f_2^2&=\frac{1}{9}\sqrt{6\cG}\,T ^{-\tfrac{3}{2}}~, & \rho^2&=\frac{\kappa^2}{\sqrt{6\cG}} T^{\tfrac{1}{2}}~,
\end{align}
and composite quantities have been defined as
\begin{align}\label{eq:kappa2-G}
	\kappa^2&=-|\partial_w \cA_+|^2+|\partial_w \cA_-|^2~,
	&
	\partial_w\cB&=\cA_+\partial_w \cA_- - \cA_-\partial_w\cA_+~,
	\nonumber\\
	\cG&=|\cA_+|^2-|\cA_-|^2+\cB+\bar{\cB}~,
	&
	T^2&=\left(\frac{1+R}{1-R}\right)^2=1+\frac{2|\partial_w\cG|^2}{3\kappa^2 \, \cG }~.
\end{align}
For the solutions of interest here $\Sigma$ has a boundary, on which the $S^2$ collapses to form a closed internal space. The differentials $\partial\cA_\pm$ have poles on the boundary of $\Sigma$. At these points, $w=r_\ell$, 5-branes emerge, with charges $(p_\ell,q_\ell)$ given by the residues of $\partial\cA_\pm$,
\begin{align}\label{eq:charges}
	\Res_{w=r_\ell} \partial_w\cA_\pm & = \frac{3}{4}\alpha^\prime (\pm q_\ell +i p_\ell)~.
\end{align}
Solutions for 5-brane junctions with 7-branes were constructed in \cite{DHoker:2017zwj}; they in addition have punctures with $SL(2,\ZZ)$ monodromy.
Solutions with O7 planes ($D_4$ 7-branes) were constructed in \cite{Uhlemann:2019lge} as $\ZZ_2$ quotients. The procedures discussed below generalize this construction.

The $SL(2,\ZZ)$ duality transformations of the Type IIB supergravity fields are induced by $SU(1,1)\otimes \CC$ transformations of the functions $\cA_\pm$. In the conventions of \cite{DHoker:2017zwj},
\begin{align}\label{eq:cApm-SU11}
	\begin{pmatrix} \cA_+\\ \cA_- \end{pmatrix} &\rightarrow
	\begin{pmatrix} u & -v \\ -\bar v & \bar u \end{pmatrix}
	\begin{pmatrix} \cA_+\\ \cA_- \end{pmatrix}
	+\begin{pmatrix} a_+\\ \bar a_+\end{pmatrix}
\end{align}
where $|u|^2-|v|^2=1$. The $SU(1,1)$ part induces $SL(2,\RR)$ transformations on the Type IIB  supergravity fields in which the axion-dilaton transforms as
\begin{align}\label{eq:tau-SU11}
	\tau&\rightarrow \frac{a\tau+b}{c\tau+d}=\frac{-i(u+\bar u-v-\bar v)\tau -u+\bar u-v+\bar v}{(u-\bar u-v+\bar v)\tau-i(u+\bar u+v+\bar v)}~,
\end{align}
while the shifts by $a_+$ induce gauge transformations on the two-form fields.
The expression for $u,v$ in terms of the $SL(2,\ZZ)$ parameters $a,b,c,d$ is 
\begin{align}\label{eq:uv-abcd}
	u&=\frac{1}{2}(a+ib-ic+d)~, & v&=\frac{1}{2}(-a+ib+ic+d)~.
\end{align}

\subsection{$T_N/\ZZ_3$ and $E_6$ 7-branes}
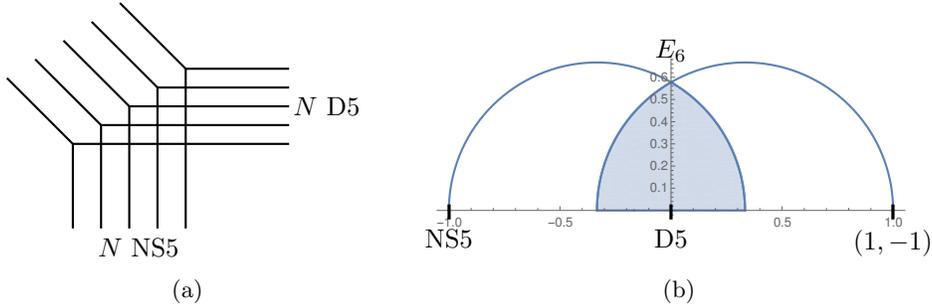
\begin{figure}
	\subfigure[][]{\label{fig:TN-web}
		\begin{tikzpicture}[xscale=-0.5,yscale=0.5]
			\draw[thick] (-4,0.75) -- (-0.5,0.75) -- (-0.5,-3);
			\draw[thick] (-0.5,0.75) -- +(1.75,1.75);
			
			\draw[thick] (-4,0.25) -- (0.25,0.25) -- (0.25,-3);
			\draw[thick] (0.25,0.25) -- +(1.75,1.75);
			
			\draw[thick] (-4,-0.25) -- (1.0,-0.25) -- (1.0,-3);
			\draw[thick] (1.0,-0.25) -- +(1.75,1.75);
			
			\draw[thick] (-4,-0.75) -- (1.75,-0.75) -- (1.75,-3);
			\draw[thick] (1.75,-0.75) -- +(1.75,1.75);
			
			\draw[thick] (-4,1.25) -- (-1.25,1.25) -- (-1.25,-3);
			\draw[thick] (-1.25,1.25) -- +(1.75,1.75);
			
			\node at (0,-3.5) {\small $N$ NS5};
			\node at (-5,0.25) {\small $N$ D5};
		\end{tikzpicture}
	}\hskip 4mm
	\subfigure[][]{\label{fig:TN-quot-H}
	\begin{tikzpicture}
		\node at (0,0.85) {\includegraphics[width=0.38\linewidth]{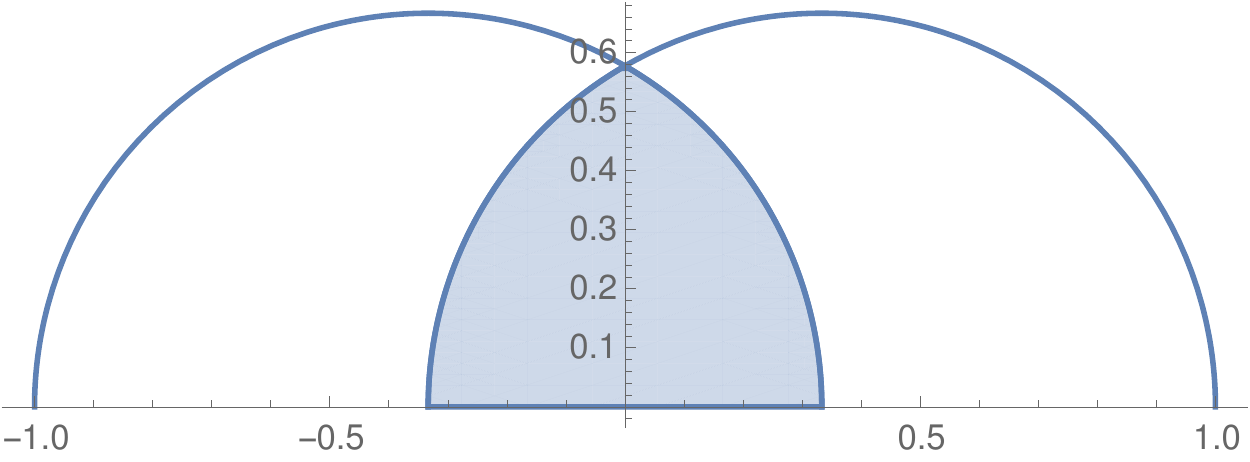}};
		\draw [very thick] (2.95,0.05) -- (2.95,-0.15) node [anchor=north] {\small $(1,-1)$};
		\draw [very thick] (0,0.05) -- (0,-0.15) node [anchor=north] {\small D5};
		\draw [very thick] (-2.95,0.05) -- (-2.95,-0.15) node [anchor=north] {\small NS5};
		\node at (0,2.1) {\small $E_6$};
	\end{tikzpicture}
	}
	\caption{Left: brane web for the $T_N$ theories. Right: Supergravity solution (\ref{eq:Apm-TN}) on the upper half plane with complex coordinate $w$. A fundamental domain for the $\ZZ_3$ quotient by (\ref{eq:w-transf-TN}) is shaded. The two circle segments connecting the $\ZZ_3$ fixed point $w=i/\sqrt{3}$ to $w=\pm1/3$ are identified by the $\ZZ_3$ quotient. This leads to a branch cut with the monodromy of an $E_6$ 7-brane and the appropriate conical deficit at $w=i/\sqrt{3}$.}
	\end{figure}

We start with the $T_N$ theories, fig.~\ref{fig:TN-web}.
The functions $\cA_\pm$ and $\cG$ specifying the supergravity solution associated with the UV fixed point were spelled out explicitly in \cite{Uhlemann:2020bek}. Taking the Riemann surface $\Sigma$ as the upper half plane with complex coordinate $w$,
\begin{align}\label{eq:Apm-TN}
	\cA_\pm&=\frac{3}{8\pi}N \left[\mp\ln\frac{w-1}{w+1}-i\ln\frac{2w}{w+1}\right]~,
	&
	\cG&=\frac{9}{8\pi^2}N^2D\left(\frac{2w}{w+1}\right)~,
\end{align}
where $D$ is the Bloch-Wigner dilogarithm function
\begin{align}
	D(z)&=\Im\left(\Li_2(z)+\ln(1-z)\ln|z|\right)~.
\end{align}
In all-in-going convention the 5-brane charges at the junction should add to zero. We have a D5-brane pole in $\partial \cA_\pm$ corresponding to charge $(-N,0)$ at $w=0$,
an NS5-brane pole corresponding to charge $(0,N)$ at $w=1$, and a $(N,-N)$ 5-brane pole at $w=-1$.

This solution has a $\ZZ_3$ symmetry, which consists of an $SL(2,\RR)$ transformation of the upper half plane permuting the poles and a compensating Type IIB $SL(2,\ZZ)$ duality transformation.
The $SL(2,\RR)$ transformation of the upper half plane cyclically permuting the poles is given by
\begin{align}\label{eq:w-transf-TN}
	w&\rightarrow  w^\prime=\frac{1+w}{1-3w}~.
\end{align}
This transformation has a fixed point at $w=i/\sqrt{3}$.
It leaves $\cG$ invariant, since $D(1-1/z)=D(z)$, but not $\cA_\pm$. 
The compensating $SU(1,1)\times \CC$ transformation is given by (\ref{eq:cApm-SU11}) with
\begin{align}\label{eq:SU11-TN}
	u^{}_{E_6}&=i-\frac{1}{2}~, & v^{}_{E_6}&=-\frac{1}{2}~,
	&
	a_+&=\frac{3N}{8}(1+i)~.
\end{align}
The functions $\cA_\pm$ are then invariant under the combination of the two transformations, i.e.
\begin{align}
	\begin{pmatrix} \cA_+(w)\\ \cA_-(w)\end{pmatrix}&\rightarrow 
	\begin{pmatrix}
		 u^{}_{E_6} \cA_+(w^\prime)-v^{}_{E_6}\cA_-(w^\prime)+a_+
		 \\
		 -\bar v^{}_{E_6} \cA_+(w^\prime)+\bar u^{}_{E_6}\cA_-(w^\prime)+\bar a_+
	\end{pmatrix}.
\end{align}

The $\ZZ_3$ S-fold amounts to quotienting the upper half plane by the $\ZZ_3$ action (\ref{eq:w-transf-TN}).
This leaves the fundamental domain shown in fig.~\ref{fig:TN-quot-H}.
At the $\ZZ_3$ fixed point, $w=i/\sqrt{3}$, the total angle is reduced from $2\pi$ to $2\pi/3$, producing a conical deficit of $4\pi/3$. The monodromy across the branch cut is given by the $SU(1,1)$ transformation (\ref{eq:SU11-TN}).
The mapping to $SL(2,\ZZ)$ via (\ref{eq:tau-SU11}), (\ref{eq:uv-abcd}) leads to the element
\begin{align}
	K_{E_6}=(ST)^{4}&=
	\begin{pmatrix}
		0& 1\\ -1 & -1
	\end{pmatrix}~.
\end{align}
This matches the transformation for an $E_6$ 7-brane e.g.\ in table 3 of \cite{Kim:2021fxx}, and the deficit angle matches as well. Since the fixed point $w=i/\sqrt{3}$ is invariant under the $SU(1,1)$ transformation (\ref{eq:SU11-TN}), the axion-dilaton at that point also takes the required value. So the quotient produced a 7-brane of type $E_6$ at the point $w=i/\sqrt{3}$.
The 7-brane is captured in the solution with its full backreaction, i.e.\ deficit angle and monodromy.

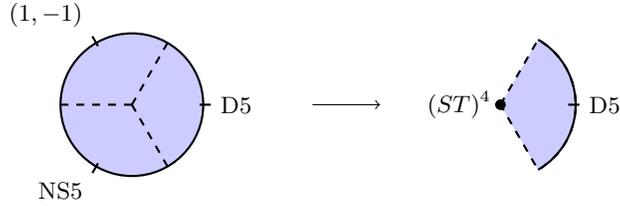
\begin{figure}
	\begin{tikzpicture}
		\draw[thick,fill=blue!20] (0,0) circle (27pt);

		\draw[thick,dashed] (0,0) -- ({cos(-60)},{sin(-60)});
		\draw[thick,dashed] (0,0) -- ({cos(180)},{sin(180)});
		\draw[thick,dashed] (0,0) -- ({cos(60)},{sin(60)});
		
		\draw[thick] (0.9,0) -- (1.05,0) node [anchor=west] {\footnotesize D5};
		\draw[thick] ({0.9*cos(120)},{0.9*sin(120)}) -- ({1.05*cos(120)},{1.05*sin(120)}) node [anchor=south east] {\footnotesize $(1,-1)$};
		\draw[thick] ({0.9*cos(240)},{0.9*sin(240)}) -- ({1.05*cos(240)},{1.05*sin(240)})  node [anchor=north east] {\footnotesize NS5};
		
		\draw[->] (2.4,0) -- (3.3,0);
		\begin{scope}[xshift=4.9cm]
		\draw[thick,dashed,fill=blue!20] (0,0) --  (60:1) arc (60:-60:1) -- cycle;
		\draw[thick] (60:1) arc (60:-60:1);
		
		\draw[thick] (0.9,0) -- (1.05,0) node [anchor=west] {\footnotesize D5};	
		\draw [fill=black] (0,0) circle (1.8pt) node [anchor=east] {\footnotesize $(ST)^4$};
		\end{scope}
	\end{tikzpicture}

\caption{The $T_N$ solution (\ref{eq:Apm-TN}) on the unit disc with coordinate $z$ defined in (\ref{eq:z-coord-TN}), with the quotient realizing the $\ZZ_3$ S-fold. The dashed lines on the right are identified. \label{fig:TN-quot-disc}}
\end{figure}

\medskip

\textbf{S-fold on the disc:}
The discussion above can be phrased naturally on the disc. To this end, we map the upper half plane to the unit disc with coordinate $z$ via
\begin{align}\label{eq:z-coord-TN}
	z&=\frac{i-\sqrt{3}w}{i+\sqrt{3}w}~.
\end{align}
This maps the poles in $\partial_w\cA_\pm(w)$ at $w=\lbrace 0,\pm 1\rbrace$ to poles in $\partial_z \cA_\pm(z)$ at cubic roots of unity, as shown on the left in fig.~\ref{fig:TN-quot-disc}.
The transformation (\ref{eq:w-transf-TN}) becomes a rotation by 120 degrees,
\begin{align}
	z&\rightarrow e^{\frac{2 \pi i}{3}}z~.
\end{align}
Taking the quotient by (\ref{eq:w-transf-TN}) combined with the $SU(1,1)$ transformation (\ref{eq:SU11-TN}) amounts to cutting the disc into three slices and keeping one, while identifying the edges.
We can for example, with the convention $\arg z\in (-\pi,\pi)$, restrict $z$ to 
\begin{align}
	|\arg(z)|\leq \frac{\pi}{3}~.
\end{align}
This is shown in fig.~\ref{fig:TN-quot-disc}. The deficit angle $4\pi/3$ is now manifest. We may map this back to a full disc with a single 7-brane branch cut by introducing a new coordinate $z=v^{1/3}$.

\subsection{$+_{N,N}/\ZZ_4$  and $E_7$ 7-branes}

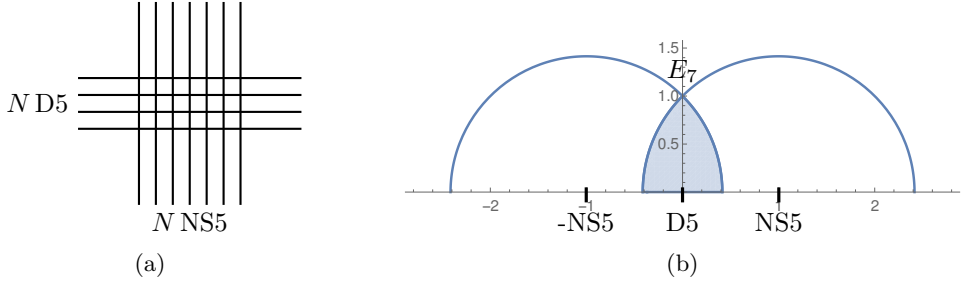
\begin{figure}
	\subfigure[][]{\label{fig:plus-web}
		\begin{tikzpicture}[scale=0.45]
			\foreach \i in {-1.5,-1.0,-0.5,0,0.5,1,1.5}{
				\draw[thick] (\i,-3) -- (\i,3);
			}
			\foreach \j in {-0.75,-0.25,0.25,0.75}{
				\draw[thick] (3.3,\j) -- (-3.3,\j);
			}			
			
			\node at (0,-3.5) {\small $N$\,NS5};
			\node at (-4.5,0) {\small $N$\,D5};
			
		\end{tikzpicture}
	}\hskip 10mm
	\subfigure[][]{\label{fig:plus-quot-H}
	\begin{tikzpicture}
		\node at (0,0.85) {\includegraphics[width=0.45\linewidth]{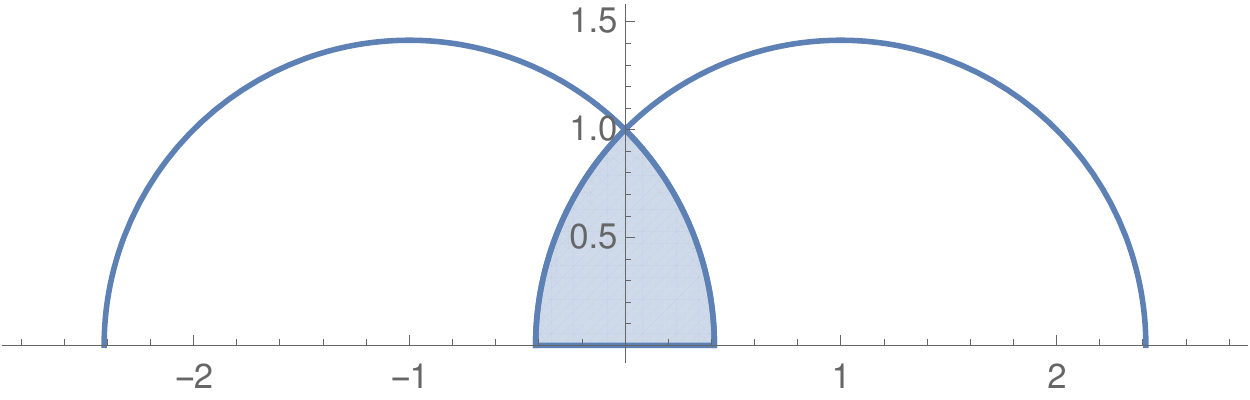}};
		\draw [very thick] (1.28,0.05) -- (1.28,-0.15) node [anchor=north] {\small NS5};
		\draw [very thick] (0,0.05) -- (0,-0.15) node [anchor=north] {\small D5};
		\draw [very thick] (-1.28,0.05) -- (-1.28,-0.15) node [anchor=north] {\small -NS5};
		\node at (0,1.6) {\small $E_7$};
	\end{tikzpicture}
	}
	\caption{Left: brane web for the  $+_{N,N}$ theories. Right: Supergravity solution (\ref{eq:cA-plus-N}) on the upper half plane, with a fundamental domain for the $\ZZ_4$ quotient shaded. The two circle segments connecting the $\ZZ_4$ fixed point $w=i$ to $w=\pm (\sqrt{2}-1)$ are identified by the $\ZZ_4$ quotient.}
\end{figure}

The next example is the $+_{N,N}$ theory, fig.~\ref{fig:plus-web}.
The functions $\cA_\pm$, $\cG$ for the $+_{N,M}$ theories were spelled out explicitly in \cite{Uhlemann:2020bek}. Here we specialize to $N=M$ and use an $SL(2,\RR)$ transformation to rewrite them in a convenient form, such that
\begin{align}\label{eq:cA-plus-N}
	\cA_\pm&=\frac{3N}{8\pi}\left[-i\ln w\pm \ln \frac{ w+1}{ w-1}\right]~, 
&
	\cG&=\frac{9N^2}{8\pi^2}\left[D\left(\frac{w+1}{ w-1}\right)+D\left(\frac{1-w}{1+w}\right)\right]~.
\end{align}
We have a pole representing D5-branes with charge $(-N,0)$ at $w=0$, a pole representing NS5-branes with charge $(0,-N)$ at $w=1$, D5-branes with charge $(N,0)$ at $w=\infty$ and NS5-branes with charge $(0,N)$ at $w=-1$.

The brane web has a $\ZZ_4$ symmetry which is also realized in the supergravity solution.
As before, we start with an $SL(2,\RR)$ transformation of the upper half plane which cyclically permutes the poles -- in the present case generating a $\ZZ_4$. The transformation is
\begin{align}\label{eq:plus-Z4}
	w&\rightarrow w^\prime=\frac{1+w}{1-w}~.
\end{align}
Combining this transformation of the upper half plane with an $SU(1,1)\times\CC$ transformation (\ref{eq:cApm-SU11}) with
\begin{align}\label{eq:plus-SU11}
	u_{E_7}^{}&=i~, & v_{E_7}^{}&=0~, & a_+=\frac{3iN}{8}~,
\end{align}
generates a symmetry of the functions $\cA_\pm$. They are invariant under
\begin{align}
	\begin{pmatrix} \cA_+(w)\\ \cA_-(w)\end{pmatrix}&\rightarrow 
	\begin{pmatrix}
		u^{}_{E_7} \cA_+(w^\prime)-v^{}_{E_7}\cA_-(w^\prime)+a_+
		\\
		-\bar v^{}_{E_7} \cA_+(w^\prime)+\bar u^{}_{E_7}\cA_-(w^\prime)+\bar a_+
	\end{pmatrix}.
\end{align}
Translating the $SU(1,1)$ transformation (\ref{eq:plus-SU11}) to $SL(2,\RR)$ via (\ref{eq:uv-abcd}) leads to the element
\begin{align}
	K_{E_7}=S^3&=\begin{pmatrix} 0 & 1 \\ -1 & 0\end{pmatrix}~.
\end{align}
Taking a quotient with respect to the $\ZZ_4$ symmetry generated by the combined transformations produces the S-fold.
The fundamental domain is shown in fig.~\ref{fig:plus-quot-H} on the upper half plane.
The total angle at the fixed point is reduced from $2\pi$ to $\frac{\pi}{2}$, so the deficit angle is $\frac{3\pi}{2}$. The deficit angle and monodromy match the expectations for an $E_7$ 7-brane.

\medskip

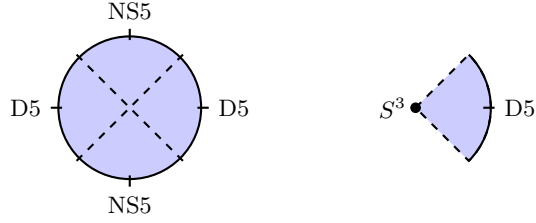
\begin{figure}
	\begin{tikzpicture}
		\draw[thick,fill=blue!20] (0,0) circle (27pt);
		
		\draw[thick,dashed] ({-cos(45)},{-sin(45)}) -- ({cos(45)},{sin(45)});
		\draw[thick,dashed] ({-cos(45)},{sin(45)}) -- ({cos(45)},-{sin(45)});
		
		\draw[thick] (0.9,0) -- (1.05,0) node [anchor=west] {\footnotesize D5};
		\draw[thick] (-0.9,0) -- (-1.05,0) node [anchor=east] {\footnotesize D5};
		\draw[thick] (0,0.9) -- (0,1.05) node [anchor=south] {\footnotesize NS5};
		\draw[thick] (0,-0.9) -- (0,-1.05) node [anchor=north] {\footnotesize NS5};
		
		\begin{scope}[xshift=3.8cm]
			\draw[thick,dashed,fill=blue!20] (0,0) --  (45:1) arc (45:-45:1) -- cycle;
			\draw[thick] (45:1) arc (45:-45:1);
			
			\draw[thick] (0.9,0) -- (1.05,0) node [anchor=west] {\footnotesize D5};	
			
			\draw [fill=black] (0,0) circle (1.8pt) node [anchor=east] {\footnotesize $S^3$};
		\end{scope}
	\end{tikzpicture}

	\caption{The $+_{N,N}$ solution (\ref{eq:cA-plus-N}) on the unit disc with coordinate $z$ in (\ref{eq:z-coord-plus}). The $\ZZ_4$ quotient amounts to keeping one slice, with the dashed lines on the right identified. \label{fig:plus-quot-disc}}
\end{figure}

{\bf S-fold on the disc:}
As for the $T_N$ solution before, a simpler geometric picture can be obtained by mapping the solution to the unit disc. We use
\begin{align}\label{eq:z-coord-plus}
	z&=\frac{i-w}{i+w}~.
\end{align}
The poles are mapped to $z\in\lbrace\pm 1,\pm i\rbrace$. 
This leads to the picture in fig.~\ref{fig:plus-quot-disc}.
The transformation in (\ref{eq:plus-Z4}) now corresponds to $z\rightarrow iz$. The quotient corresponds to decomposing the disc into 4 slices and keeping one of them with the edges identified.
The fundamental domain can be taken as $|\arg(z)|\leq \frac{\pi}{4}$.
The total angle $\pi/2$ at the fixed point is now manifest.

\subsection{$\pslash_N/\ZZ_6$ and $E_8$ 7-branes}

\begin{figure}
	\subfigure[][]{\label{fig:pslash-web}
		\begin{tikzpicture}[yscale=-0.5,xscale=-0.5]
			\foreach \i in {1/2,1,3/2,2,5/2}{
				\draw (-4,0.5*\i) circle (1.5pt)  -- (2.5-\i,0.5*\i) -- +(1.5,1.5);
				\draw (2.5-\i,0.5*\i) -- (2.5-\i,-3);
				\draw[fill] (2.5-\i,-3) circle (1.5pt);
				\draw[fill] (-4,0.5*\i) circle (1.5pt);
				\draw[fill] (4-\i,1.5+0.5*\i) circle (1.5pt);
				\begin{scope}[yshift=-0.25cm]
					\draw (3.5,0.5-0.5*\i) -- (-3+\i,0.5-0.5*\i) -- +(-1.5,-1.5);
					\draw (-3+\i,0.5-0.5*\i) -- (-3+\i,3.5);
					\draw[fill] (3.5,0.5-0.5*\i) circle (1.5pt);
					\draw[fill] (-3+\i,3.5) circle (1.5pt);
					\draw[fill] (-4.5+\i,-1-0.5*\i) circle (1.5pt);
				\end{scope}
			}
		\end{tikzpicture}
	}\hskip 4mm
	\subfigure[][]{\label{fig:pslash-fund}
	\begin{tikzpicture}
		\node at (0,0.8) {\includegraphics[width=0.45\linewidth]{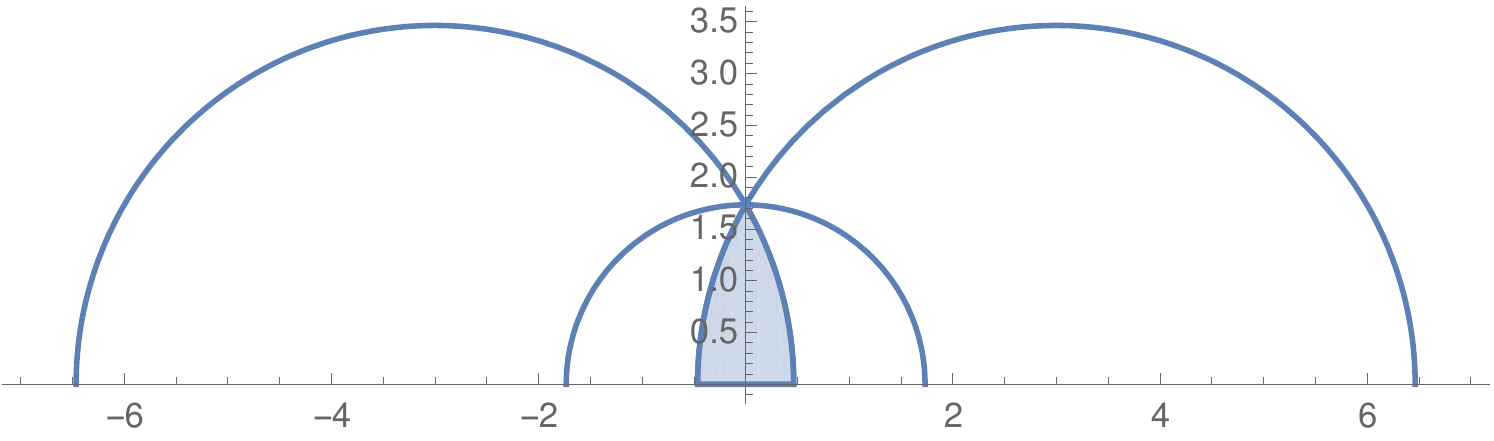}};
		\draw [very thick] (1.54,0.05) -- (1.54,-0.15) node [anchor=north] {\small };
		\draw [very thick] (0,0.05) -- (0,-0.15) node [anchor=north] {\small D5};
		\draw [very thick] (-1.54,0.05) -- (-1.54,-0.15) node [anchor=north] {\small};
		
		\draw [very thick] (0.51,0.05) -- (0.51,-0.15) node [anchor=north] {\small };
		\draw [very thick] (-0.51,0.05) -- (-0.51,-0.15) node [anchor=north] {\small };
		\node at (0,1.3) {\footnotesize $E_8$};
	\end{tikzpicture}
	}
	\caption{Left: brane web for the $\pslash_N$ theories. Right: Supergravity solution (\ref{eq:cA-pslash}) on the upper half plane, with a fundamental domain for the $\ZZ_6$ quotient shaded.}
\end{figure}
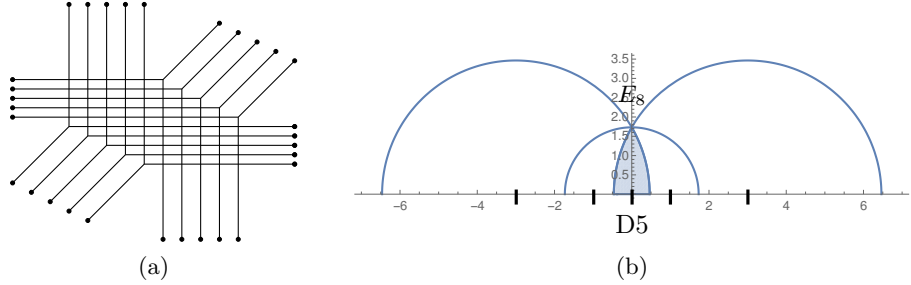

The last example is based on the $\pslash_N$ theory of \cite{Bergman:2018hin}. The brane web is shown in fig.~\ref{fig:pslash-web} and the supergravity dual was only given implicitly in \cite{Bergman:2018hin}. Here we take the functions $\cA_\pm$ as
\begin{align}\label{eq:cA-pslash}
	\cA_\pm&=\frac{3N}{8\pi}\left[(\pm 1+i)\ln\frac{w+3}{w-1}-i\ln w\pm\ln\frac{w+1}{w-3}\right]\,.
\end{align}
The function $\cG$ can be constructed explicitly following appendix C of \cite{Uhlemann:2020bek}; we will not need it here.
We have a D5 pole with charge $(-N,0)$ at $w=0$, a $(-N,-N)$ pole at $w=1$, an NS5 pole with charge $(0,-N)$ at $w=3$, a D5 pole with charge $(N,0)$ at $w=\infty$, an $(N,N)$ pole at $w=-3$ and an NS5 pole with charge $(0,N)$ at $w=-1$.

The brane web has a $\ZZ_6$ symmetry which is reflected in the supergravity solutions.
Cyclic permutations of the poles are generated by
\begin{align}\label{eq:w-transf-pslash}
	w&\rightarrow w^\prime=\frac{3w+3}{3-w}~.
\end{align}
The compensating $SU(1,1)\times\CC$ transformation is given by (\ref{eq:cApm-SU11}) with
\begin{align}
	u^{}_{E_8}&=i+\frac{1}{2}~, & v^{}_{E_8}&=\frac{1}{2}~, &
	a_+&=-\frac{3iN}{8}~.
\end{align}
The combined symmetry transformation leaving $\cA_\pm$ invariant is
\begin{align}
	\begin{pmatrix} \cA_+(w)\\ \cA_-(w)\end{pmatrix}&\rightarrow 
	\begin{pmatrix}
		u^{}_{E_8} \cA_+(w^\prime)-v^{}_{E_8}\cA_-(w^\prime)+a_+
		\\
		-\bar v^{}_{E_8} \cA_+(w^\prime)+\bar u^{}_{E_8}\cA_-(w^\prime)+\bar a_+
	\end{pmatrix}.
\end{align}
Taking a quotient of the solution with respect to this symmetry realizes the $\ZZ_6$ quotient of the brane web. 
A fundamental domain in the supergravity solution is shown in fig.~\ref{fig:pslash-fund}.
The total angle at the fixed point $w=i\sqrt{3}$ is reduced to $\pi/3$, with a deficit angle $5\pi/3$.
Translating the $SU(1,1)$ transformation to $SL(2,\RR)$ leads to 
\begin{align}
	K_{E_8}=(TS)^5&=\begin{pmatrix}0&1\\-1&1\end{pmatrix}~.
\end{align}
We have thus realized the deficit angle and monodromy of an $E_8$ 7-brane at the fixed point.

\medskip

\begin{figure}
	\begin{tikzpicture}
		\draw[thick,fill=blue!20] (0,0) circle (27pt);
		
		\foreach \i in {30,90,150,210,270,330} \draw[thick,dashed] (0,0) -- ({cos(\i)},{sin(\i)});
		\foreach \i in {0,60,120,180,240,300} \draw[thick,dashed] ({0.9*cos(\i)},{0.9*sin(\i)}) -- ({1.1*cos(\i)},{1.1*sin(\i)});
		
		\node at (1.3,0) {\footnotesize D5};
		\node at (-1.3,0) {\footnotesize D5};		
		\node at ({-1.3*cos(240)},{-1.3*sin(240)})  {\footnotesize (-1,-1)};
		\node at ({1.3*cos(240)},{-1.3*sin(240)})  {\footnotesize NS5};
		\node at ({1.3*cos(240)},{1.3*sin(240)})  {\footnotesize (1,1)};
		\node at ({-1.3*cos(240)},{1.3*sin(240)})  {\footnotesize NS5};
		
		\draw[->] (2.4,0) -- (3.3,0);
		\begin{scope}[xshift=4.9cm]
			\draw[thick,dashed,fill=blue!20] (0,0) --  (30:1) arc (30:-30:1) -- cycle;
			\draw[thick] (30:1) arc (30:-30:1);
			
			\draw[thick] (0.9,0) -- (1.05,0) node [anchor=west] {\footnotesize D5};	
			\draw [fill=black] (0,0) circle (1.8pt) node [anchor=east] {\footnotesize $(ST)^5$};
		\end{scope}
	\end{tikzpicture}
	
	\caption{Quotient $\pslash_N/\ZZ_6$ on the disc with coordinate $z$. The dashed lines on the right are identified. \label{fig:pslash-quot-disc}}
\end{figure}
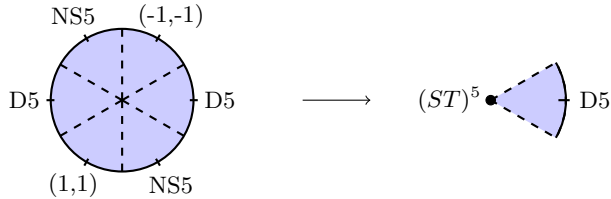

{\bf S-fold on the disc:}
To again phrase the discussion on the disc we map the upper half plane to the unit disc via
\begin{align}
	z&=\frac{i\sqrt{3}-w}{i\sqrt{3}+w}~.
\end{align}
This maps the poles in $\partial_w\cA_\pm(w)$ at $w=\lbrace 0,\pm 1,\pm 3,\infty\rbrace$ to poles in $\partial_z \cA_\pm(z)$ at sextic roots of unity, as shown on the left in fig.~\ref{fig:pslash-quot-disc}.
The transformation (\ref{eq:w-transf-pslash}) becomes a rotation by 60 degrees,
\begin{align}
	z&\rightarrow e^{\frac{2 \pi i}{6}}z~.
\end{align}
The $\ZZ_6$ quotient amounts to cutting the disc into six slices and keeping one, with the edges identified.
We can for example restrict $z$ to $|\arg(z)|\leq \frac{\pi}{6}$.
This is shown in fig.~\ref{fig:TN-quot-disc}.

\subsection{S-folds leading to $H_{0,1,2}$ 7-branes}

So far we discussed quotients by $\ZZ_n$ symmetries to realize 7-branes of type $E_{6,7,8}$.
In this section we implement the generalized quotients of sec.~\ref{sec:H-S-folds} to realize 7-branes of type $H_{0,1,2}$. For simplicity we work directly with the representation of the solutions on the unit disc.

\smallskip

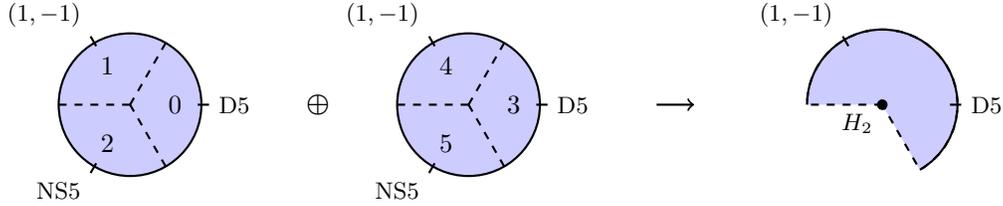
\begin{figure}
		\begin{tikzpicture}
		\begin{scope}[xshift=-4.5cm]
				\draw[thick,fill=blue!20] (0,0) circle (27pt);
				
				\draw[thick,dashed] (0,0) -- ({cos(-60)},{sin(-60)});
				\draw[thick,dashed] (0,0) -- ({cos(180)},{sin(180)});
				\draw[thick,dashed] (0,0) -- ({cos(60)},{sin(60)});
				
				\draw[thick] (0.9,0) -- (1.05,0) node [anchor=west] {\footnotesize D5};
				\draw[thick] ({0.9*cos(120)},{0.9*sin(120)}) -- ({1.05*cos(120)},{1.05*sin(120)}) node [anchor=south east] {\footnotesize $(1,-1)$};
				\draw[thick] ({0.9*cos(240)},{0.9*sin(240)}) -- ({1.05*cos(240)},{1.05*sin(240)})  node [anchor=north east] {\footnotesize NS5};

				\node at ({0.6*cos(0)},{0.6*sin(0)}) {\small $0$};
				\node at ({0.6*cos(120)},{0.6*sin(120)}) {\small $1$};
				\node at ({0.6*cos(240)},{0.6*sin(240)}) {\small $2$};
		\end{scope}
	
		\node at (-2,0) {{\boldmath{$\oplus$}}};
	
		\draw[thick,fill=blue!20] (0,0) circle (27pt);
		
		\draw[thick,dashed] (0,0) -- ({cos(-60)},{sin(-60)});
		\draw[thick,dashed] (0,0) -- ({cos(180)},{sin(180)});
		\draw[thick,dashed] (0,0) -- ({cos(60)},{sin(60)});
		
		\draw[thick] (0.9,0) -- (1.05,0) node [anchor=west] {\footnotesize D5};
		\draw[thick] ({0.9*cos(120)},{0.9*sin(120)}) -- ({1.05*cos(120)},{1.05*sin(120)}) node [anchor=south east] {\footnotesize $(1,-1)$};
		\draw[thick] ({0.9*cos(240)},{0.9*sin(240)}) -- ({1.05*cos(240)},{1.05*sin(240)})  node [anchor=north east] {\footnotesize NS5};
		
		\node at ({0.6*cos(0)},{0.6*sin(0)}) {\small $3$};
		\node at ({0.6*cos(120)},{0.6*sin(120)}) {\small $4$};
		\node at ({0.6*cos(240)},{0.6*sin(240)}) {\small $5$};

		\draw[->,thick] (2.5,0) -- (3,0);
		\begin{scope}[xshift=5.5cm]
			\draw[thick,dashed,fill=blue!20] (0,0) --  (180:1) arc (180:-60:1) -- cycle;
			\draw[thick] (180:1) arc (180:-60:1);
			
			\draw[thick] (0.9,0) -- (1.05,0) node [anchor=west] {\footnotesize D5};	
			\draw[thick] ({0.9*cos(120)},{0.9*sin(120)}) -- ({1.05*cos(120)},{1.05*sin(120)}) node [anchor=south east] {\footnotesize $(1,-1)$};
			
			\draw [fill=black] (0,0) circle (1.8pt) node [anchor=north east] {\footnotesize ${H_2}$};
		\end{scope}
	\end{tikzpicture}
	\caption{Two copies of $T_N$ solutions, with each disc decomposed into 3 slices, labeled by $k=0,\ldots 5$. The $\ZZ_3$ action corresponds to moving the slices according to $k\rightarrow k+2 \mod 6$. This is a rotation by 240 degrees, with the refinement that the slices get bumped to the other disc if they cross the dividing line between the first and last slice. The $\ZZ_3$ quotient amounts to keeping two out of three slices of one $T_N$ solution, as shown on the right. The dashed lines are identified and we have created an $H_2$ 7-brane at the center.\label{fig:H2-disc}}
\end{figure}

{\bf \boldmath{$(T_N)^{2}/\ZZ_3$} and $H_2$ 7-branes:}
For the $T_N$ theory, the standard $\ZZ_3$ quotient was realized by slicing the associated supergravity solution on the disc into three equal pieces and keeping one, with the edges identified (fig.~\ref{fig:TN-quot-disc}).
To realize the `2/3 quotient', we start with two copies of the $T_N$ solution on the disc, so that we have two discs. We label the slices of the first disc by $k=0,1,2$ and those of the second by $k=3,4,5$. We then define a new $\ZZ_3$ action by $k\rightarrow k+2 \mod 6$. Pictorially, each slice gets moved two spots over, as in a 240 degree rotation, with the refinement that if a slice crosses the dividing line between the third and first slice on its disc it gets bumped to the other disc. This action links the two copies of the disc non-trivially. The quotient with respect to this $\ZZ_3$ transformation identifies all points in a $\ZZ_3$ orbit. A fundamental domain is given by two slices of one of the discs. 
This result can be described as starting with one $T_N$ solution, slicing up the disc as before, but keeping two slices, as shown in fig.~\ref{fig:H2-disc}. In that sense we may describe this operation as a `2/3 quotient' of a single $T_N$ theory.
We call this the $(T_N)^{\,2/3}$ theory.
The total angle around the fixed point at the center is $4\pi/3$, leaving a deficit angle of $2\pi/3$. This is the expected value for an $H_2$ 7-brane.
The monodromy at the fixed point is $(K_{E_6})^2$, which is conjugate to that of an $H_2$ 7-brane,
\begin{align}
	K_{H_2}&=K_{[0,1]}(K_{E_6})^2K_{[0,1]}^{-1}~.
\end{align}
The $H_2$ 7-brane monodromy can be realized directly by transforming the seed solution by a $K_{[0,1]}$ $SL(2,\ZZ)$ transformation.

\smallskip

\begin{figure}
	\subfigure[][]{\label{fig:H1-disc}
			\begin{tikzpicture}
			\draw[thick,fill=blue!20] (0,0) circle (27pt);
			
			\draw[thick,dashed] ({-cos(45)},{-sin(45)}) -- ({cos(45)},{sin(45)});
			\draw[thick,dashed] ({-cos(45)},{sin(45)}) -- ({cos(45)},-{sin(45)});
			
			\draw[thick] (0.9,0) -- (1.05,0) node [anchor=west] {\footnotesize D5};
			\draw[thick] (-0.9,0) -- (-1.05,0) node [anchor=east] {\footnotesize D5};
			\draw[thick] (0,0.9) -- (0,1.05) node [anchor=south] {\footnotesize NS5};
			\draw[thick] (0,-0.9) -- (0,-1.05) node [anchor=north] {\footnotesize NS5};
			
			\draw[->] (1.9,0) -- (2.4,0);
			\begin{scope}[xshift=4.4cm]
				\draw[thick,dashed,fill=blue!20] (0,0) --  (225:1) arc (225:-45:1) -- cycle;
				\draw[thick] (225:1) arc (225:-45:1);
				
				\draw[thick] (0.9,0) -- (1.05,0) node [anchor=west] {\footnotesize D5};
				\draw[thick] (-0.9,0) -- (-1.05,0) node [anchor=east] {\footnotesize D5};
				\draw[thick] (0,0.9) -- (0,1.05) node [anchor=south] {\footnotesize NS5};
				
				\draw [fill=black] (0,0) circle (1.8pt) node [anchor=north, yshift=-1mm] {\footnotesize $H_1$};
			\end{scope}
		\end{tikzpicture}
	}
\hfill
	\subfigure[l][]{\label{fig:H0-disc}
			\begin{tikzpicture}
			\draw[thick,fill=blue!20] (0,0) circle (27pt);
			
			\foreach \i in {30,90,150,210,270,330} \draw[thick,dashed] (0,0) -- ({cos(\i)},{sin(\i)});
			\foreach \i in {0,60,120,180,240,300} \draw[thick,dashed] ({0.9*cos(\i)},{0.9*sin(\i)}) -- ({1.1*cos(\i)},{1.1*sin(\i)});
			
			\node at (1.3,0) {\footnotesize D5};
			\node at (-1.3,0) {\footnotesize D5};		
			\node at ({-1.3*cos(240)},{-1.3*sin(240)})  {\footnotesize (-1,-1)};
			\node at ({1.3*cos(240)},{-1.3*sin(240)})  {\footnotesize NS5};
			\node at ({1.3*cos(240)},{1.3*sin(240)})  {\footnotesize (1,1)};
			\node at ({-1.3*cos(240)},{1.3*sin(240)})  {\footnotesize NS5};
			
			\draw[->] (1.9,0) -- (2.4,0);
			\begin{scope}[xshift=4.4cm]
				\draw[thick,dashed,fill=blue!20] (0,0) --  (-30:1) arc (-30:270:1) -- cycle;
				\draw[thick] (-30:1) arc (-30:270:1);
				\foreach \i in {0,60,120,180,240} \draw[thick,dashed] ({0.9*cos(\i)},{0.9*sin(\i)}) -- ({1.1*cos(\i)},{1.1*sin(\i)});
				
				\draw[thick] (0.9,0) -- (1.05,0) node [anchor=west] {\footnotesize D5};
				\node at (-1.3,0) {\footnotesize D5};		
				\node at ({-1.3*cos(240)},{-1.3*sin(240)})  {\footnotesize (-1,-1)};
				\node at ({1.3*cos(240)},{1.3*sin(240)})  {\footnotesize (1,1)};
				\node at ({1.3*cos(240)},{-1.3*sin(240)})  {\footnotesize NS5};
					
				\draw [fill=black] (0,0) circle (1.8pt) node [anchor=north west,yshift=-2mm] {\footnotesize $H_0$};
			\end{scope}
		\end{tikzpicture}
	}
	\caption{The analogous constructions to fig.~\ref{fig:H2-disc}, for $+_{N,N}$ with one out of four slices taken out and the cuts identified in \ref{fig:H1-disc}, and for $\pslash_N$ with one out of six slices taken out in \ref{fig:H0-disc}.}
\end{figure}
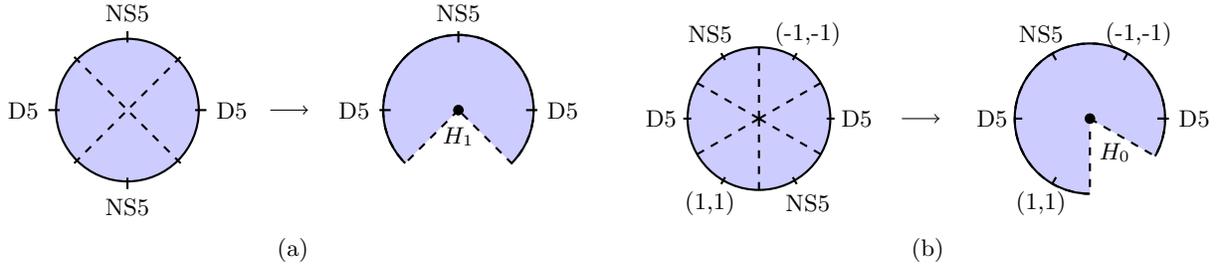

{\bf \boldmath{$(+_{N,N})^{3}/\ZZ_4$} and $H_1$ 7-branes:}
A generalized $\ZZ_4$ quotient of the $+_{N,N}$ solution can be realized accordingly. We start with $n-1=3$ copies of the solution on the disc and decompose each disc into $n=4$ slices. We label the slices by $k=0,\ldots ,(n-1)n-1$. A new $\ZZ_4$ action is now defined by reshuffling the slices according to $k\rightarrow k+4 \mod 12$. This corresponds to a 90 degree rotation with the refinement that if a slice crosses the dividing line between the first and last slice it gets bumped to the next copy of the disc.
The result of taking a quotient with respect to this $\ZZ_4$ action can be described as decomposing one disc into four slices and keeping three of them, as shown in fig.~\ref{fig:H1-disc}.
We will refer to this theory as $(+_{N,N})^{3/4}$. The procedure leads to a deficit angle of $\pi/2$ at the fixed point at the center, in line with the expectation for an $H_1$ 7-brane. The monodromy is $(K_{E_7})^3$, which is the monodromy of an $H_1$ 7-brane.

\smallskip
{\bf \boldmath{$(\pslash_N)^5/\ZZ_6$} and $H_0$ 7-branes:} 
The last example is the generalized $\ZZ_6$ quotient of the $\pslash_N$ solution. Here we start with $n-1=5$ copies of the solution on the disc, decompose each disc into $n=6$ slices, and label the slices by $k=0,\ldots,n(n-1)-1$. The $\ZZ_6$ action is then defined by moving the slices according to $k\rightarrow k+5 \mod n(n-1)$.
This can be understood as a rotation by 60 degrees combined with bumping slices crossing the dividing line between the first and last slice on a disc to the next disc. The $\ZZ_6$ quotient with respect to this action amounts to keeping 5 out of 6 slices of one disc, as shown in fig.~\ref{fig:H0-disc}. We refer to this theory as $(\pslash_N)^{5/6}$. The deficit angle at the fixed point at the center is $\pi/3$. Combined with the monodromy and the value of the axio-dilaton, we find an $H_0$ 7-brane at the center.

\section{Free energies, central charges, defect operators}\label{sec:observables}

The holographic duals give access to many observables in the large-$N$ limit. Since the S-fold theories do not generically admit a low-energy non-abelian gauge theory description, this is particularly useful.
We discuss a sample of observables in this section. They will be connected to matrix models in sec.~\ref{sec:matrix-models}.

\smallskip
{\bf Free energies:}
The $S^5$ free energies are of particular interest as they are conjecturally monotonic along renormalization group flows.
The free energies are related in a simple way between the parent and quotient theories at large $N$. Since they are computed holographically from the on-shell action, with an $SL(2,\ZZ)$ invariant Type IIB Lagrangian, a $\ZZ_n$ quotient simply divides the on-shell action by $n$. For the $T_N$, $+_{N,N}$ and $\pslash_N$ theories the free energies obtained holographically were matched to field theory calculations in \cite{Uhlemann:2019ypp}. From these results we conclude for the S-fold theories
\begin{align}\label{eq:FS5-quotients}
	F_{S^5}[T_N/\ZZ_3]&=-\frac{9}{8\pi^2}\zeta(3)N^4~,
	&
	F_{S^5}[+_{N,N}/\ZZ_4]&=-\frac{189}{64\pi^2}\zeta(3)N^4~,
	\nonumber\\
	F_{S^5}[\pslash_N/\ZZ_6]&=-\frac{63}{4\pi^2}\zeta(3)N^4~.
\end{align}
For the $(T_N)^{2/3}$, $(+_{N,N})^{3/4}$ and $(\pslash_N)^{5/6}$ theories the free energies can be obtained similarly from those of the parent theories, leading to
\begin{align}
	F_{S^5}[(T_N)^{\,2/3}]&=2F_{S^5}[T_N/\ZZ_3]~,
	&
	F_{S^5}[(+_{N,N})^{3/4}]&=3F_{S^5}[+_{N,N}/\ZZ_4]~,
	\nonumber\\
	F_{S^5}[(\pslash_N)^{5/6}]&=5F_{S^5}[\pslash_N/\ZZ_6]~.
\end{align}
A general expression for arbitrary 5-brane junctions was derived as a form of F-minimization in \cite{Fluder:2020pym}. From this expression the free energy can be obtained also  for more general quotient theories.

\smallskip
{\bf Central charges, topologically twisted indices:}
The stress tensor 2-point function central charge $C_T$ and the (universal) topologically twisted index, computed from the partition function on ${\Sigma_{\mathfrak{g}_1}\times \Sigma_{\mathfrak{g}_2}\times S^1}$, are related to the sphere free energy through the relations
\begin{align}\label{eq:CT-TTI}
	C_T&=-\frac{640}{\pi^2}F_{S^5}~,
	&
	\ln \mathcal Z_{\Sigma_1\times \Sigma_2\times S^1}&=-\frac{8}{9}(1-\mathfrak{g}_1)(1-\mathfrak{g}_2)F_{S^5}~.
\end{align}
The former were matched to field theory calculations in \cite{Fluder:2018chf, Uhlemann:2019ypp}, the latter in \cite{Fluder:2019szh}. 
The central charge $C_T$ is related to the squashed sphere partition function (see e.g.\ \cite{Chang:2017cdx}).
Holographically, the universal relations (\ref{eq:CT-TTI}) follow from the existence of consistent truncations to 6d gauged supergravity \cite{Hong:2018amk,Malek:2018zcz}.
They hold in the parent theories, where they can be reproduced from gauge theory calculations, and they are inherited by the quotient theories.

\smallskip
{\bf Line operators:}
Line operators play an interesting role for 5d theories with holographic duals in Type IIB. They can be used to locally link the brane webs to the supergravity solutions and, where gauge theory deformations are available, to the matrix models arising from supersymmetric localization  \cite{Uhlemann:2020bek}. For theories with quiver gauge theory descriptions with all nodes balanced the connection was further made explicit in \cite{Legramandi:2021uds,Fatemiabhari:2022kpv}. 

D3-branes which are pointlike in the plane of the 5-brane web and describe line operators in the field theory can be placed in every compact face of the 5-brane web \cite{Assel:2012nf}. In the large-$N$ limit, where the external 5-brane charges are large, this leads to a 2-parameter family of line operators. In quiver gauge theory terms these are Wilson loops in antisymmetric representations associated with individual gauge nodes, labeled by the node and the rank of the representation. In the holographic duals there is a $\tfrac{1}{2}$-BPS probe D3-brane wrapping $\rm AdS_2\times S^2$ at each point of the Riemann surface $\Sigma$ \cite{Uhlemann:2020bek}. It carries F1 and D1 charges given by
\begin{align}
	N_{\rm F1}+i N_{\rm D1}&=\frac{4}{3}\left(\cA_++\bar\cA_-\right)~.
\end{align}
They specify, respectively, the vertical and horizontal coordinates of the face in the brane web (the gauge node and rank of the representation).
The expectation value is given by
\begin{align}
	\ln\left\langle W_{\wedge}\right\rangle&=-\frac{2}{3}T_{\rm D3}\Vol_{AdS_2}\Vol_{S^2}\cG~.
\end{align}
S-folding reduces $\Sigma$ to a (fractional) quotient. 
The standard $\ZZ_n$ S-folds thus retain a fraction $1/n$ of the line operators associated with D3-branes of the seed theory, while the fractional quotients retain a fraction $(n-1)/n$. The expectation values are unchanged.

\medskip
{\bf Surface operators:} 
The study of 3d defect operators represented by D3-branes ending on the 5-brane web was initiated for simple 5-brane webs in \cite{Gaiotto:2014ina}. In the supergravity duals, conformal defects are represented by D3-branes wrapping $\rm AdS_4$, localized at a point on $\Sigma$ \cite{Gutperle:2020rty}.
For each of the $T_N$, $+_{N,N}$ and $\pslash_N$ theories, there is one conformal D3-brane wrapping $\rm AdS_4$. It is localized on $\Sigma$ at the fixed points of the $Z_n$ actions discussed in sec.~\ref{sec:duals} where the 7-brane emerges upon S-folding. The defect D3-branes carry no fluxes and are invariant under $SL(2,\ZZ)$ and the $\ZZ_n$ symmetry of the solutions. The D3-brane defect operators should therefore be part of the S-fold theories.

\medskip
{\bf Universal short multiplets:} Two families of universal operators in short multiplets, common to all 5d SCFTs with holographic duals in Type IIB, were identified in \cite{Gutperle:2018wuk}. In the notation of \cite{Cordova:2016emh}, they correspond to $B_2$ multiplets with scalar primaries $[0,0]_{\Delta=3\ell+3}^{(2\ell)}$ and $A_2$ multiplets with scalar primaries $[0,0]_{\Delta=4+2\ell}^{(2\ell)}$, where $\ell$ is a non-negative integer.
They arise from spin-2 fluctuations,
\begin{align}
	ds^2_{\rm AdS_6}&\rightarrow ds^2_{\rm AdS_6}+h^{[tt]}_{\mu\nu}(x) Y_{\ell m}(S^2)\phi_{\ell m}(w,\bar w) dx^\mu dx^\nu
	&
	\square_{AdS_6}h^{[tt]}_{\mu\nu}&= ( M^2 -2) h^{[tt]}_{\mu\nu}~,
\end{align} 
around the metric (\ref{eqn:ansatz}), where $x^\mu$ denotes the coordinates on $\rm AdS_6$ and $h^{[tt]}_{\mu\nu}$ is a transverse-traceless fluctuation on $\rm AdS_6$.
The universal solutions are 
\begin{align}
	\phi^{(0)}_{\ell m}&=\cG^\ell~, & M^2&=3\ell(3\ell+5)~,
	\nonumber\\
	\phi^{(1)}_{\ell m}+i\phi^{(2)}_{\ell m}&=\cG^\ell(\cA_+-\bar\cA_-)~, & M^2&=(3\ell+1)(3\ell+6)~.
\end{align}
The modes $\phi^{(0)}_{\ell m}$ are invariant under $SL(2,\ZZ)$, while
$(\phi^{(1)}_{\ell m},\phi^{(2)}_{\ell m})$ form a doublet. However, being constructed out of $\cA_\pm$, all modes are invariant under the $\ZZ_n$ symmetries discussed in sec.~\ref{sec:duals} and describe universal spin-2 operators in the S-fold theories. 

\begin{figure}
	\begin{tikzpicture}
		
		\draw[thick,fill=blue!20] (0,0) circle (27pt);
		
		\draw[red,thick] (0,0) -- (1,0);
		\draw[red,thick] (0,0) -- ({cos(120)},{sin(120)});
		\draw[red,thick] (0,0) -- ({cos(-120)},{sin(-120)});
		
		\draw[dashed] (0,0) -- ({cos(-60)},{sin(-60)});
		\draw[dashed] (0,0) -- ({cos(180)},{sin(180)});
		\draw[dashed] (0,0) -- ({cos(60)},{sin(60)});
		
		\draw[thick] (0.9,0) -- (1.05,0) node [anchor=west] {\footnotesize D5};
		\draw[thick] ({0.9*cos(120)},{0.9*sin(120)}) -- ({1.05*cos(120)},{1.05*sin(120)}) node [anchor=south east] {\footnotesize $(1,-1)$};
		\draw[thick] ({0.9*cos(240)},{0.9*sin(240)}) -- ({1.05*cos(240)},{1.05*sin(240)})  node [anchor=north east] {\footnotesize NS5};

		\begin{scope}[xshift=3.8cm]
			\draw[thick,dashed,fill=blue!20] (0,0) --  (60:1) arc (60:-60:1) -- cycle;
			\draw[thick] (60:1) arc (60:-60:1);
			
			\draw[red,thick] (0,0) -- (1,0);
			
			\draw[thick] (0.9,0) -- (1.05,0) node [anchor=west] {\footnotesize D5};	
			\draw [fill=black] (0,0) circle (1.8pt) node [anchor=east] {\footnotesize $(ST)^4$};
		\end{scope}
	\end{tikzpicture}
	\hskip 10mm
	\begin{tikzpicture}[scale=0.85]
		\foreach \i in {-3,...,3}{
			\draw (0.0,0.17*\i) -- +(2.5,0) [fill] circle (1.5pt);
		}
		\draw[fill=gray!20] (0,0) circle (0.65);
		\foreach \i in {-3,-1,1,3}{
			\draw[fill] (0,0.17*\i) circle (1.5pt);
			\draw[thick,dashed] (0,0.17*\i) -- +(-1.9,0);
		}
		\node at (-2.2,3*0.17) {\footnotesize $A^5$};
		\node at (-2.29,1*0.17) {\footnotesize $B$};
		\node at (-2.29,-1*0.17) {\footnotesize $C$};
		\node at (-2.29,-3*0.17) {\footnotesize $B$};
		
		\draw[red,thick]   (0,3*0.17) .. controls  (0.5,1) and (2,1) .. (2.5,0.17*3);
		\draw[blue,thick] (0,1*0.17) -- (0.3,0.08) -- (0.3,-0.08) -- (0,-1*0.17);
		\draw[blue,thick]   (0.3,0.08) .. controls  (0.5,0.5) and (2,0.5) .. (2.5,0.17*1);
		\draw[blue,thick]   (0.3,-0.08) .. controls  (0.5,-0.5) and (2,-0.5) .. (2.5,-0.17*1);
		
		\node at (0,-1.4) {};
	\end{tikzpicture}
	
	\caption{Left: string junction representing the trifundamental $\Delta=3N$ operator in $T_N$. Center: strings connecting the D5-brane pole to the $E_6$ 7-brane in the S-fold. Right: after resolving the $E_6$ 7-brane.\label{fig:TN-string}}
\end{figure}
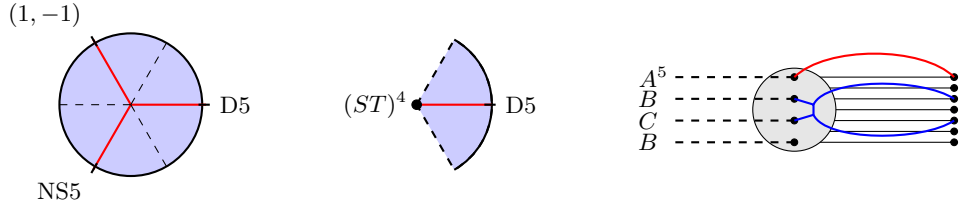

\medskip
{\bf Stringy operators:}
An interesting part of the spectrum are BPS operators represented by string junctions in the brane webs and supergravity solutions, with scaling dimensions of order $N$ in the large-$N$ limit \cite{Bergman:2018hin}.
For the $T_N$ theory the junction of D1, F1 and $(1,1)$ string segments on the left in fig.~\ref{fig:TN-string} represents an operator in the trifundamental of $SU(N)^3$ with scaling dimension $\Delta=\frac{3}{2}(N-1)$. 
Some components of this operator were identified in the gauge theory in \cite{Bergman:2018hin}. 

For the $T_N/\ZZ_3$ S-fold we have stringy operators connecting the 5-brane pole on the boundary of $\Sigma$ to the 7-brane emerging at the $\ZZ_3$ fixed point. This is shown in fig.~\ref{fig:TN-string} for single and double string operators connecting the D5-brane pole to (a resolution of) the $E_6$ 7-brane.
For $N=3,4,5$ the $T_N/\ZZ_3$ theories have gauge theory descriptions, but for general $N$ we do not know a gauge theory description to relate to. The scaling dimensions of these operators can nevertheless be determined from the supergravity duals.
For the two operators shown in the figure this yields
\begin{align}
	\Delta_{\rm F1}&=N~, & \Delta_{{2\rm F1}}&=2N~.
\end{align}
The flavor symmetry algebra is $\mathfrak{g}_F= \mathfrak{e}_6\times \mathfrak{su}(N)$, and at large $N$ we do not expect further enhancement, unlike for $N=3$. The $\mathfrak{g}_F$ representations  can be determined from the resolution on the right in fig.~\ref{fig:TN-string}.
Operators for other S-folds can be discussed analogously.

\smallskip
{\bf Descendant theories:} All the observables above can be discussed analogously for more general S-folds. An interesting example are S-folds of the $B_N$ theories described in  \cite{Eckhard:2020jyr}, which can be obtained from the $T_N$ theories by mass deformations. The brane webs for these theories have external 5-branes with charges $(1,N-1)$, $(N-2,-1)$ and $-(N-1,N-2)$, and have a similar $\ZZ_3$ symmetry as the $T_N$ theories. They are higher-rank generalizations of the $E_0$ Seiberg theory, with no gauge theory descriptions to begin with. The supergravity solutions are, to leading order at large $N$, identical to those for the $T_N$ theories, since the 5-brane charges only differ at $\mathcal O(1)$. As a result, observables like the leading-order free energy are also identical at large $N$, while for example the spectrum of stringy operators differs.

\smallskip

Further aspects of the S-folds which may be studied along similar lines include compactifications \cite{Legramandi:2021aqv}, mass deformations \cite{Akhond:2022awd}, and connections of the Type IIB solutions to M-theory \cite{Kaidi:2018zkx}.

\section{Matrix models}\label{sec:matrix-models}

For 5d SCFTs with gauge theory deformations, matrix models encoding BPS observables can be derived using supersymmetric localization. For S-fold theories we in general have no gauge theory descriptions.
However, based on the form of the matrix models associated with the seed theories, which in our examples have gauge theory deformations, and the prepotentials derived from the brane webs, we will propose matrix models which capture at least certain observables at large~$N$.

The matrix models derived from supersymmetric localization for the seed theories take the form of an integral over the Coulomb branch, with an integrand involving the classical prepotential, one-loop determinants, and instanton contributions. The latter are expected to be suppressed at large $N$ \cite{Jafferis:2012iv,Choi:2019miv} and we will drop them.
We then define matrix models for the S-fold theories by incorporating the identifications of the Coulomb branch parameters, derived from the brane webs in sec.~\ref{sec:S-fold-review}, and substituting the effective brane web prepotentials for the S-folds.
While we do not necessarily expect these matrix models to capture all aspects of the S-fold theories, they efficiently encode at least certain observables, including the free energies, at large $N$.

The partition functions of 5d gauge theories on squashed five-spheres were computed in \cite{Kallen:2012va,Kim:2012ava,Imamura:2012efi,Lockhart:2012vp}.
For a theory with gauge group $G$ and $N_f$ hypermultiplets, the partition function can be written as 
\begin{align}
	\mathcal{Z}_{\vec{\omega}} = \frac{S'_3(0|\vec{\omega})^{{\rm rk}G}}{|\mathcal{W}|(2\pi)^{{\rm rk}G}} \int [d\lambda]e^{-\frac{2\pi}{6\omega_1\omega_2\omega_3}\kappa\, {\rm tr}\lambda^3}\times \frac{\prod_{e\in {\rm root}}S_3(i e(\lambda)|\vec{\omega})}{\prod_f\prod_{w_f\in {\rm weight}}S_3(i w_f(\lambda)+\frac{\omega_{\rm tot}}{2}|\vec{\omega})} \times Z_{\rm inst} \ ,
\end{align}
where $S_3$ is the triple sine function, $\omega_{1,2,3}$ are the squashing parameters, $\omega_{\rm tot}\equiv\frac{\omega_1+\omega_2+\omega_3}{2}$, ${\rm rk}G$ is the rank of the gauge group, and $\kappa$ is the Chern-Simons level. Here, $Z_{\rm inst}$ is the instanton contribution and all mass parameters are switched off. We shall assume that the instanton contribution is suppressed at large $N$.

The calculation of the (squashed) $S^5$ partition functions in the long quiver large-$N$ limit relevant for the gauge theories considered here has been performed in \cite{Uhlemann:2019ypp}, and we will adapt it to the S-fold theories below. Assuming that the integral is dominated by large $\lambda_i$ in the large-$N$ limit, the partition function becomes
\begin{align}\label{eq:Z-gen-1}
	\mathcal{Z}_{\vec{\omega}} &= \int [d\lambda] e^{-\frac{1}{\omega_1\omega_2\omega_3}\mathcal{F}_{\vec{\omega}}} \nonumber  \\
	\mathcal{F}_{\vec{\omega}}&= \frac{\pi}{3}\kappa \, {\rm tr}\lambda^3 + \sum_{e\in {\rm root}}F_V(e(\lambda)) + \sum_f\sum_{w_f\in {\rm root}}F_H(w_f(\lambda)) \ ,
\end{align}
where
\begin{align}\label{eq:Z-gen-2}
	F_V(x) &\equiv -\frac{1}{2}\omega_1\omega_2\omega_3\big[\log S_3(ix|\vec\omega)+\log S_3(-ix|\vec\omega)\big] \approx \frac{\pi}{6}|x|^3 - \frac{\omega^2_{\rm tot}+\omega_1\omega_2+\omega_2\omega_3+\omega_3\omega_1}{12}\pi|x|~, \nonumber \\
	F_H(x) &= \omega_1\omega_2\omega_3\log S_3(ix+\frac{\omega_{\rm tot}}{2}|\vec\omega) \approx -\frac{\pi}{6}|x|^3 - \frac{\omega_1^2+\omega_2^2+\omega_3^2}{24}\pi |x| \ .
\end{align}
The function $\mathcal{F}_{\vec\omega}$ coincides with the effective prepotential evaluated on the squashed $S^5$, and the expressions (\ref{eq:Z-gen-1}), (\ref{eq:Z-gen-2}) will be the starting point for the examples to be discussed in the following.

\subsection{\texorpdfstring{$T_N/\mathbb{Z}_3$}{T[N]/Z[3]} and $(T_N)^{2/3}$}

We start with the $T_N/\ZZ_3$ and $(T_N)^{2/3}$ theories. 
The $T_N$ theory before any quotients has a gauge theory description as
\begin{align}\label{eq:TN-quiver}
	 [N]-SU(N-1)-SU(N-2)-\cdots - SU(3)-SU(2)-[2] \ .
\end{align}
The prepotential for the $T_N$ theory is given by
\begin{align}\label{eq:FTN}
	\mathcal{F}^{T_N}_{\vec\omega} &= \sum_{t=1}^{N-2}\sum_{\ell \neq m}^{N-t}F_V(\lambda_{\ell}^{(t)}-\lambda_{m}^{(t)}) + \sum_{t=1}^{N-3}\sum_{\ell=1}^{N-t}\sum_{m=1}^{N-t-1}F_H(\lambda_\ell^{(t)}- \lambda_m^{(t+1)}) \nonumber \\
	&+N\sum_{\ell=1}^{N-1}F_H(\lambda_\ell^{(1)})+2\sum_{\ell=1}^{2}F_H(\lambda_\ell^{(N-2)}) \ .
\end{align}
In this expression, we used the scalars $\lambda_\ell^{(t)}$ for the $t^{\rm th}$  gauge node $SU(N-t)$ in the orthogonal basis with constraint $\sum_{\ell =1}^{N-t} \lambda_\ell^{(t)}=0$.

We now perform the $\mathbb{Z}_3$ folding of the $T_N$ theory which was investigated in \cite{Acharya:2021jsp,Kim:2021fxx}. Under the $\mathbb{Z}_3$ quotient, the vector multiplet scalars in the $T_N$ theory are identified as \cite{Kim:2021fxx}
\begin{align}\label{eq:TN-identification}
	\phi^{(t)}_i = \phi^{(i)}_{N-t-i} \ ,
\end{align}
where $\phi_i^{(t)}$ denotes the scalars in the $SU(\cdot)$ Dynkin basis with the relation $\phi_i^{(t)} = \sum_{\ell = 1}^i \lambda_\ell^{(t)}$. The effective prepotential for $T_N/\mathbb{Z}_3$ can then be obtained from the prepotential of the seed theory in (\ref{eq:FTN}) by dividing it by $1/3$ and implementing the identifications as
\begin{align}\label{eq:cF-TN-Z3}
	\mathcal{F}_{\vec\omega}^{T_N/\mathbb{Z}_3} = \frac{1}{3}\mathcal{F}_{\vec\omega}^{T_N} + \sum_{t=1}^{N-2}\sum_{i=1}^{N-t-1}W_i^t\left(\sum_{\ell=1}^i\lambda^{(t)}_{\ell}-\sum_{m=1}^{N-t-i}\lambda_m^{(i)}\right) \ ,
\end{align}
where we have introduced Lagrange multipliers $W^t_i$ to impose the constraint (\ref{eq:TN-identification}).

For the large-$N$ limit we introduce eigenvalue densities $\rho^{(t)}(\lambda)$ for each gauge node, and, following \cite{Uhlemann:2019ypp}, combine them into a single function of two variables, $\hat\rho(z,x)$, where $x$ is defined by $\lambda=(N-1)\omega_{\rm tot}x$ and $z=t/(N-1)\in[0,1]$ labels the gauge nodes. The expression for the prepotential then becomes (\cite[eq.~(4.8)]{Uhlemann:2019ypp}, adapted to match the orientation of the quiver in (\ref{eq:TN-quiver}))
\begin{align}\label{eqn:cFTN-6}
	\cF_{T_N}  &=
	N^2\int dz\,dx\,dy\,\cL
	+N^5\int dx\, \hat\rho(0,x)\left[F_H(x)-\frac{1}{2}\int dy\,\hat\rho(z,y)F_H(x-y)\right]\,,
\end{align}
where
\begin{align}\label{eq:cL}
	\cL&=\hat N(z)^2\hat\rho(z,x)\hat\rho(z,y)F_0(x-y\big)
	-\frac{1}{2}\partial_z \big(\hat N(z)\hat\rho(z,x)\big)\partial_z\big(\hat N(z)\hat\rho(z,y)\big)F_H\big(x-y\big)~.
\end{align}
The rank function is $\hat N(z)=N (1-z)$ and $F_0(x)\equiv(F_V(x)-F_H(x))/\omega_{\rm tot}^2$. 

In the planar limit the matrix integral is dominated by a saddle point, and the leading-order free energy is determined by the effective prepotential evaluated on that saddle point. 
If the saddle point for the $T_N$ theory satisfies the constraints in (\ref{eq:TN-identification}), then replacing the effective prepotential of the $T_N$ theory in (\ref{eq:FTN}) by that for the $T_N/\ZZ_3$ theory in (\ref{eq:cF-TN-Z3}) does not affect the saddle point equations. The free energy then simply picks up a factor $1/3$.
 
The saddle point eigenvalue densities for the $T_N$ theory are given in \cite[eq.~(4.9)]{Uhlemann:2019ypp}, which, with $z\rightarrow 1-z$ to match the orientation of the quiver in (\ref{eq:TN-quiver}), reads
\begin{align}\label{eq:TN-saddle}
	\hat\rho_s(z,x)&=\frac{\sin (\pi  z)}{1-z}\frac{1}{\cosh \left(2\pi x\right)-\cos (\pi  z)}~.
\end{align}
The constraints enforced by the Lagrange multipliers in (\ref{eq:cF-TN-Z3}) translate to
\begin{align}\label{eq:TN-constraint}
	\phi_{i}^{(t)}/(N\omega_{\rm tot})&=\hat N(z_t)\int_{-\infty}^{x_0} dx\,x\, \hat\rho(z_t,x)\stackrel{!}{=}\hat N(z_i)\int_{-\infty}^{x_1}dx\,x\, \hat\rho(z_i,x)=\phi_{N-t-i}^{(i)}/(N\omega_{\rm tot})~,
\end{align}
where $z_t=t/(N-1)$, $z_i=i/(N-1)$ and $x_0$ and $x_1$ are determined from
\begin{align}
	i&=\hat N(z_t)\int_{-\infty}^{x_0}dx\,\hat\rho(z_t,x)\,, 
	& N-t-i&=\hat N(z_i)\int_{-\infty}^{x_1} dx\,\hat\rho(z_i,x)\,.
\end{align}
That is, $x_0$ and $x_1$ are determined such that the integrals on the left and right hand sides in (\ref{eq:TN-constraint}) sum up the first $i$ and $N-t-i$ eigenvalues at the $t^{\rm th}$ and $i^{\rm th}$ gauge node, respectively.
From (2.32), (2.33) of \cite{Uhlemann:2020bek} (with $z\rightarrow 1-z$),
\begin{align}
	x_0&=K\left(z_t,\frac{i}{\hat N(z_t)}\right)~,
	&
	x_1&=K\left(z_i,\frac{N-t-i}{\hat N(z_i)}\right)~,
\end{align}
where
\begin{align}
	K(z,y)&=\frac{1}{2\pi}\ln\left[\sin(\pi y(1-z))\csc(\pi(1-y)(1-y))\right]~.
\end{align}
With these definitions one can verify that the constraint in (\ref{eq:TN-constraint}) is indeed satisfied by the family of saddle point densities. The computation of the free energy therefore follows as in \cite{Uhlemann:2019ypp}, up to an overall factor $1/3$ resulting from (\ref{eq:cF-TN-Z3}). This matches the holographic prediction in (\ref{eq:FS5-quotients}).

The discussion straightforwardly extends to the $(T_N)^{2/3}$ theory, for which the effective prepotential follows from (\ref{eq:F-fractional}). Since we have shown that the saddle point of the $T_N$ theory is compatible with the $\ZZ_3$ action on the Coulomb branch parameters derived from the brane web, the free energy calculation at large $N$ differs from that for the $T_N$ theory only by the appropriate overall factors.

The central charge $C_T$ is obtained from the free energy on squashed spheres by expanding in the squashing parameters \cite{Chang:2017cdx}. This was shown to lead to the relation in (\ref{eq:CT-TTI}) for the parent theories of the S-folds in \cite{Uhlemann:2019ypp}. Since our matrix models for the S-fold theories do not change the dependence on the squashing parameters compared to the parent theories, they reproduce the same relation. 

\subsection{\texorpdfstring{$+_{N,N}/\ZZ_4$}{plus[N,N]/Z[4]} and $(+_{N,N})^{3/4}$}
The discussion for the $+_{N,N}/\ZZ_4$ and $(+_{N,N})^{3/4}$ theories follows analogously.
The $+_{N,N}$ theory before the quotient has a gauge theory description as
\begin{align}
	[N]-SU(N) -\ldots - SU(N)-[N]
\end{align}
with a total of $N-1$ gauge nodes.
The function $\mathcal{F}_{\vec\omega}$ can be written as
\begin{align}\label{eq:+NN}
	\mathcal{F}^{+_{N,N}}_{\vec{\omega}}&=\sum_{t=1}^{N-1}\sum_{\ell\neq m}^{N-1}F_V(\lambda_{\ell}^{(t)}-\lambda_{m}^{(t)}) + \sum_{t=1}^{N-2}\sum_{\ell=1}^{N-1}\sum_{m=1}^{N-1}F_H(\lambda_\ell^{(t)}- \lambda_m^{(t+1)}) \nonumber \\
	&+N\sum_{\ell=1}^{N-1}F_H(\lambda_\ell^{(1)})+N\sum_{\ell=1}^{N-1}F_H(\lambda_\ell^{(N-1)}) \ .
\end{align}
We can use this function to compute the effective prepotential of the $\mathbb{Z}_4$ quotient of the $+_{N,N}$ theory. The $\mathbb{Z}_4$ quotient identifies the vector multiplet scalars $\phi_i^{(t)}$ in Dynkin basis  as
\begin{align}\label{eq:+NN-identification}
	\phi_i^{(t)} = \phi_{N-t}^{(i)} \ , 
\end{align}
where $\phi_i^{(t)} = \sum_{\ell = 1}^i \lambda_\ell^{(t)}$. As discussed in sec.~\ref{sec:E-S-folds}, the effective prepotential for the $\mathbb{Z}_4$ quotient can be obtained as a $1/4$ fraction of the original prepotential in (\ref{eq:+NN}), with the identification of the scalar fields as (\ref{eq:+NN-identification}). We thus have
\begin{align}\label{eq:constrained-F-plus}
	\mathcal{F}^{+_{N,N}/\mathbb{Z}_4}_{\vec{\omega}} = \frac{1}{4}\mathcal{F}^{+_{N,N}}_{\vec{\omega}} + \sum_{i=1}^{N-1}\sum_{t=1}^{N-1}W_i^t\left(\sum_{\ell=1}^i \lambda_\ell^{(t)} - \sum_{m=1}^{N-t}\lambda_m^{(i)}\right) \ ,
\end{align}
where $W^t_i$ denotes the Lagrange multipliers for the constraint (\ref{eq:+NN-identification}).

The free energy for the unquotiented $+_{N,M}$ theory was calculated in \cite{Uhlemann:2019ypp}. The saddle point eigenvalue density is given by \cite[(4.4)]{Uhlemann:2019ypp},
\begin{align}\label{eq:rho-sol-plus}
	\hat\rho_s(z,x)&=\frac{4\sin (\pi  z) \cosh \left(2 \pi x\right) }{\cosh \left(4 \pi x\right)-\cos (2 \pi  z)}~.
\end{align}
Here $z=t/(M-1)$, $\lambda = (M-1)\omega_{\rm tot} x$. The density for $\lambda$ is obtained from $\rho(z,\lambda)d\lambda=\hat\rho(z,x)dx$.
This family of saddle point eigenvalue densities satisfies the constraint in (\ref{eq:+NN-identification}). To see this, we first translate (\ref{eq:+NN-identification}) to a statement on the eigenvalue densities. It demands
\begin{align}\label{eq:plus-constraint}
	\phi_i^{(t)}&=N\int_{-\infty}^{\lambda_0} d\lambda\,\lambda\, \rho(t/N,\lambda)\stackrel{!}{=}N\int_{-\infty}^{\lambda_1}d\lambda\,\lambda\, \rho(i/N,\lambda)=\phi_{N-t}^{(i)}~,
\end{align}
where $\lambda_0$ and $\lambda_1$ are determined from
\begin{align}
	\frac{i}{N}&=\int_{-\infty}^{\lambda_0}d\lambda\,\rho(z,\lambda)\,, 
	& 1-\frac{t}{N}&=\int_{-\infty}^{\lambda_1} d\lambda\,\rho(i/N,\lambda)\,.
\end{align}
$\lambda_0$ and $\lambda_1$ are determined such that the integrals on the left and right hand sides in (\ref{eq:plus-constraint}) sum up the first $i$ and $N-t$ eigenvalues at the $t^{\rm th}$ and $i^{\rm th}$ gauge node, respectively. 
Using the integral given in \cite[(2.21)]{Uhlemann:2020bek}, with $\lambda_{0/1}\equiv M\omega_{\rm tot} x_{0/1}$, we find
\begin{align}
	\frac{i}{N}&=\frac{1}{2}+\frac{1}{\pi}\tan^{-1}\left(\sinh(2\pi x_0)\csc(\pi t/N)\right)\,,
	\nonumber\\
	\frac{t}{N}&=\frac{1}{2}-\frac{1}{\pi}\tan^{-1}\left(\sinh(2\pi x_1)\csc(\pi i/N)\right)\,.
\end{align}
This can be solved for $x_0$, $x_1$ in terms of $t$ and $i$. Upon replacing $\lambda_0$ and $\lambda_1$ in (\ref{eq:plus-constraint}) using these expressions, evaluating the constraint becomes straightforward. The result is that it is satisfied.

The free energy computation with the constraint (\ref{eq:plus-constraint}) therefore proceeds in parallel to \cite{Uhlemann:2019ypp}. The Lagrange multiplier terms in (\ref{eq:constrained-F-plus}) do not contribute on the saddle point and the free energy for the $+_{N,N}/\ZZ_4$ theory is given by that of the $+_{N,N}$ theory divided by four.
This reproduces the result quoted in (\ref{eq:FS5-quotients}).
This discussion once again extends straightforwardly to the $(+_{N,N})^{3/4}$ theory.

\subsection{\texorpdfstring{$\pslash_N/\ZZ_6$ and $(\pslash_N)^{5/6}$}{pslash[N]]/Z[6] and pslash[N]**5/6}}

We now turn to the  $\pslash_N/\ZZ_6$ and $(\pslash)^{5/6}$ theories, and compute the large-$N$ free energy starting from $\pslash_N$ before the $\ZZ_6$ quotient. The original theory has a quiver gauge theory description as 
\begin{align}
	[N]-SU(N+1)-\cdots-SU(2N-1)-SU(2N)-SU(2N-1)-\cdots -SU(N+1)-[N]
\end{align}
The effective prepotential of this theory on the squashed $S^5$ is given by
\begin{align}\label{eq:XN-pre}
	\mathcal{F}^{\pslash_N}_{\vec{\omega}}&=\sum_{t=1}^{2N-1}\sum_{\ell\neq m}^{N_t}F_V(\lambda_{\ell}^{(t)}-\lambda_{m}^{(t)}) + \sum_{t=1}^{2N-2}\sum_{\ell=1}^{N_t}\sum_{m=1}^{N_{t+1}}F_H(\lambda_\ell^{(t)}- \lambda_m^{(t+1)}) \nonumber \\
	&+N\sum_{\ell=1}^{N}F_H(\lambda_\ell^{(1)})+N\sum_{\ell=1}^{N}F_H(\lambda_\ell^{(2N-1)}) \ .
\end{align}
The $\mathbb{Z}_6$ quotient identifies the vector multiplet scalars as
\begin{align}\label{eq:XN-identification}
	\tilde\phi^{(t)}_i =\tilde{\phi}^{(i)}_{N+i-t} \quad {\rm with} \quad 
	\tilde\phi^{(t)}_i \equiv \left\{ \begin{array}{ll}\phi^{(t)}_i \ \ \text{for} \ t \le N \\
		\phi^{(t)}_{N+i-t} \ \ \text{for} \ t > N\end{array}\right. \ .
\end{align}
Then the prepotential of the $\pslash_N/\ZZ_6$ theory is given by the original prepotential (\ref{eq:XN-pre}) divided by six together with the relations (\ref{eq:XN-identification}) enforced by Lagrange multipliers. We thus find
\begin{align}\label{eq:pslash-Lagrange-multipliers}
	\mathcal{F}^{\pslash_N/\mathbb{Z}_6}_{\vec{\omega}} = \frac{1}{6}\mathcal{F}^{\pslash_N}_{\vec{\omega}} + \sum_{t=1}^{2N-1}\sum_{i=1}^{N_t}W_i^t\left(\tilde\phi^{(t)}_i - \tilde{\phi}^i_{N+i-t}\right) \ ,
\end{align}
where $W_i^t$ is the Lagrange multiplier for the constraint (\ref{eq:XN-identification}).

The saddle point for this theory is given by \cite[(4.46)]{Uhlemann:2019ypp}
\begin{align}\label{eq:pslash-saddle}
	\hat\rho_s&=
	\frac{N}{\hat N(z)}\frac{1-2 \csch^2(2\pi x+i\pi z)}{\sqrt{3}+2 \coth(2\pi x+i\pi z)}\sqrt{\frac{\sqrt{3} \tanh(2\pi x+i\pi z)+2}{\sqrt{3} \tanh(2\pi x+i\pi z)-2}}
	+\mathrm{c.c.}
\end{align}
where $\hat N(z)=2-|2z-1|$.
To verify that this saddle point satisfies the constraint enforced by the Lagrange multipliers in (\ref{eq:pslash-Lagrange-multipliers}), we first note that the family of saddle point eigenvalue densities is invariant under $z\rightarrow 1-z$ and under $x\rightarrow -x$.
So we can assume $t<N$ and $i<N$ without loss of generality for verifying the constraint (\ref{eq:XN-identification}). 
The constraint then translates to
\begin{align}\label{eq:pslash-constraint}
	\phi_{i}^{(t)}/(2N\omega_{\rm tot})&=\hat N(z_t)\int_{-\infty}^{x_0} dx\,x\, \hat\rho(z_t,x)\stackrel{!}{=}\hat N(z_i)\int_{-\infty}^{x_1}dx\,x\, \hat\rho(z_i,x)=\phi_{N+i-t}^{(i)}/(2N\omega_{\rm tot})~,
\end{align}
where $z_t=t/(2N)$, $z_i=i/(2N)$ and $x_0$ and $x_1$ are determined from
\begin{align}
	i&=\hat N(z_t)\int_{-\infty}^{x_0}dx\,\hat\rho(z_t,x)\,, 
	& N+i-t&=\hat N(z_i)\int_{-\infty}^{x_1} dx\,\hat\rho(z_i,x)\,.
\end{align}
This constraint is indeed satisfied by the saddle point in (\ref{eq:pslash-saddle}). As in the previous examples, the computation of the free energy is therefore modified compared to the original $\pslash_N$ theory only by a factor $1/6$ for $\pslash_N/\ZZ_6$ and a factor $5/6$ for the $(\pslash_N)^{5/6}$ theories.

\let\oldaddcontentsline\addcontentsline
\renewcommand{\addcontentsline}[3]{}
\begin{acknowledgments}
	We thank Christopher Couzens, Sung-Soo Kim, Neil Lambert, Kimyeong Lee, Yi-Nan Wang and the participants of the workshop ``Higher form symmetries, defects, and boundaries in QFT" at the Technion for useful discussions.
	OB is supported in part by the Israel Science Foundation under grant No.~1254/22. The research of HK is supported by the Samsung Science and Technology Foundation under Project Number SSTF-BA2002-05 and by the National Research Foundation of Korea (NRF) grant funded by the Korea government (MSIT) (No. 2018R1D1A1B07042934).
	CFU is supported, in part, by the US Department of Energy under Grant No.~DE-SC0007859 and by the Leinweber Center for Theoretical Physics. 
	Part of this work was completed at the Aspen Center for Physics, which is supported by National Science Foundation grant PHY-1607611. 
	FA, OB, and HK would like to thank the Simons Center for Geometry and Physics, Stony Brook University for the hospitality and partial support during the final stage of this work at the workshops ``Geometry of (S)QFT'' (FA, OB, HK), ``Generalized Global Symmetries, Quantum Field Theory, and Geometry'' (FA) and ``5d N=1 SCFTs and Gauge Theories on Brane Webs'' (FA, OB).
\end{acknowledgments}
\let\addcontentsline\oldaddcontentsline

\bibliography{5d-gauged-ZN}
\end{document}